\documentclass[twocolumn,pre,aps,floatfix]{revtex4-1}

\usepackage{amsmath}
\usepackage{graphicx}
\usepackage{epstopdf}
\usepackage[usenames]{color}
\usepackage{xr}
\usepackage{placeins}
\usepackage[breaklinks=true,
            pdfborder={0 0 1},
            colorlinks=true,
            linkcolor=black,
            citecolor=blue,
            urlcolor=blue]{hyperref}
\usepackage[normalem]{ulem}
\usepackage{rotating}
\usepackage{verbatim}
\usepackage{bibentry}
\usepackage{xspace}

\usepackage[dvipsnames]{xcolor}

\newcommand{\red}[1]{\textcolor{red}{#1}}

\def\eg{\textit{e.g.}\xspace}

\def\Dr{D_\text{r}}
\def\Gex{G} 
\def\kt{k_\text{B}T}

\def\kappaNew{\chi}

\def\vp{v_\text{p}}
\def\jin{j_\text{in}}
\def\jout{j_\text{out}}
\def\rhoc{\rho_\text{c}}

\def\rhog{\rho_\text{g}}

\def\ncrit{n_\text{crit}}
\def\nmax{n_\text{max}}
\def\taunucl{\tau_\text{nucl}}
\def\joutbar{\overline{j_\text{out}}}
\def\rsgn{r_\text{sgn}}

\usepackage{fancyhdr}
\usepackage{xcolor}

\begin{document}

\title{A classical nucleation theory description of active colloid assembly}

\author{Gabriel S. Redner}
\thanks{These authors contributed equally}
\author{Caleb G. Wagner}
\thanks{These authors contributed equally}
\author{Aparna Baskaran}
\author{Michael F. Hagan}
\email{hagan@brandeis.edu}

\affiliation{Martin Fisher School of Physics, Brandeis University,
  Waltham, MA, USA.}

\begin{abstract}
Non-aligning self-propelled particles with purely repulsive excluded volume interactions undergo athermal motility-induced phase separation into a dilute gas and a dense cluster phase. Here, we use enhanced sampling computational methods and analytic theory to examine the kinetics of formation of the dense phase. Despite the intrinsically nonequilibrium nature of the phase transition, we show that the kinetics can be described using an approach analogous to equilibrium classical nucleation theory, governed by an effective free energy of cluster formation with identifiable bulk and surface terms. The theory captures the location of the binodal, nucleation rates as a function of supersaturation, and the cluster size distributions below the binodal, while discrepancies in the metastable region reveal additional physics about the early stages of active crystal formation. The success of the theory shows that a framework similar to equilibrium thermodynamics can be obtained directly from the microdynamics of an active system, and can be used to describe the kinetics of evolution toward nonequilibrium steady states.
\end{abstract}
\maketitle

\thispagestyle{fancy}
\pagestyle{fancy}

Active fluids consisting of self-propelled units are
present in many biological systems, including the cell cytoplasm \cite{Brugues2012,
  Brugues2014,Goldstein2015},
bacterial suspensions \cite{Cisneros2011a,Dombrowski2004,Kaiser2014,Dunkel2013}, and animal flocks \cite{Attanasi2014b,Attanasi2014c,Shaw1978,Holldobler1994}. Recently,
researchers have also developed synthetic active fluids, consisting of
chemically\cite{Palacci2010a, Paxton2004, Hong2007, Jiang2010, Volpe2011, Thutupalli2011, Theurkauff2012,Palacci2013a,Palacci2014} or electrically ~\cite{Bricard2013}
propelled colloids, or monolayers of vibrated granular particles
\cite{Narayan2007, Kudrolli2008, Deseigne2010, Kumar2014}.  Being
intrinsically nonequilibrium, active fluids cannot be described by equilibrium statistical mechanics \cite{Marchetti2013,Bechinger2016} and exhibit behaviors
not possible in equilibrium systems, such as spontaneous flow \cite{Wan2008, Tailleur2009,Angelani2011,Ghosh2013,Ai2013,Giomi2011,Giomi2012,Giomi2012a,Voituriez2006,Marenduzzo2007a,Ginelli2006,Fielding2011,Marenduzzo2008,Giomi2008} and athermal phase
separation \cite{Fily2012,Redner2013,Stenhammar2013,Buttinoni2013a,Mognetti2013,Stenhammar2014b,Wysocki2014,Cates2015,marchetti2015structure,Stenhammar2015,Ni2013a}. Yet, active systems frequently evolve to well-defined
time-independent distributions, and in some cases are characterized
by equilibrium-like state variables such as temperature or pressure
\cite{Solon2015,Solon2015b,Takatori2014,marchetti2015structure,Takatori2015,Speck2015,Speck2015a,Winkler2015,Ginot2015}. While significant progress has been made toward
understanding these stationary distributions, the kinetics of
evolution toward steady-state remain poorly understood.


As in equilibrium physics, progress in active matter often stems from simplified model systems. 
One such system is composed of active Brownian
particles (ABPs): self-propelled particles which interact solely by
short-range isotropic repulsion. Despite lacking interparticle
attractions or alignment interactions, ABPs form macroscopic,
crystalline clusters \cite{Redner2013, Fily2012, Stenhammar2013,
  Bialke2014, Cates2015, Buttinoni2013a,Mognetti2013,Stenhammar2014b,Wysocki2013,Cates2015,Palacci2013a,Palacci2014,Theurkauff2012}. (This is an example of a generic instability toward density inhomogeneity, motility-induced phase separation (MIPS), which can arise when particle velocities decrease with increasing local particle density \cite{Tailleur2008,Cates2015,Stenhammar2013}.) ABP phase separation is strikingly
reminiscent of equilibrium vapor-liquid systems, with the densities of
the coexisting phases falling along a binodal, and critical-like
behavior near its apex.
As a minimal model system possessing nontrivial phase behavior, ABPs are ideal for
studying evolution toward steady-state in generic active systems. However, while the coarsening of deeply quenched ABP clusters has been studied numerically \cite{Redner2013,Stenhammar2013,Stenhammar2014b}, there is no theory for the complete kinetics of phase separation.
Moreover, while existing phenomenological descriptions of ABPs  have led to important insights about MIPS \cite{Wittkowski2014,Stenhammar2013,Speck2014,Bialke2015}, there is currently no approach to directly calculate phase behavior from the microdynamics of a particle-based model.

To overcome these limitations, we describe ABP
clustering dynamics and steady-state phase behavior
with  a theory analogous to classical nucleation
theory (CNT) for equilibrium phase separation \cite{Oxtoby1992,Agarwal2014,Yoreo2003}. Beginning
with a geometric picture of ABP interactions, we construct an
effective free energy of cluster formation which
 resembles that of  droplet nucleation in an
equilibrium liquid-vapor system. We then apply the framework of CNT to calculate
nucleation rates and determine phase behavior as functions of particle
density and propulsion velocity.


While previous descriptions of ABP phase separation are based on a functional ansatz at the level of pair correlations or the dependence of particle velocities on local density \cite{Wittkowski2014,Stenhammar2013,Speck2014,Bialke2015}, here we show how such frameworks  emerge from the kinetics of the microscopic model. In particular, our theory leads to a simple relationship between the microscopic parameters of an ABP model and the driving force for ABP phase separation (analogous to the chemical potential difference between dense and dilute phases in equilibrium phase separation).

We test the theory against simulations of ABPs, employing enhanced sampling
techniques to make systematic measurements of nucleation
rates. Despite approximations in our
microscopic model, the predicted
phase boundary matches simulation results almost quantitatively, with
no adjustable parameters. Moreover, the predicted and measured cluster
size distributions match well below the binodal, though we discuss interesting effects of nontrivial cluster geometry that lead to power law scaling in the metastable regime. The
theory qualitatively predicts the dependence of nucleation rates on
super-saturation, although significant quantitative differences are seen near
the binodal.




In the last century, CNT drove tremendous advances in the fields of equilibrium crystallization and self-assembly by relating particle-scale interactions to their macroscale assembly dynamics. The framework described here moves toward a similar capability in active materials, showing how the breaking of time-reversal symmetry at the level of individual particles controls emergent nonequilibrium assembly.

\textit{Model.}   ABPs in two dimensions
obey the overdamped Langevin equations:
\begin{align}
  \dot{\boldsymbol{r}}_i &= \boldsymbol{F}(\{\boldsymbol{r}_i\})/\xi + \vp \hat{\boldsymbol{\nu}}_i + \sqrt{2 D} \boldsymbol{\eta}_i^\text{T}
  \label{eq:abp-spatial}
  \\
  \dot{\theta}_i &= \sqrt{2 \Dr} \eta_i^\text{R}.
  \label{eq:abp-rotational}
\end{align}
Here $\boldsymbol{F}$ represents the interparticle repulsion force, $\xi$ is the drag,
$\vp$ is the magnitude of the self-propulsion velocity, and
$\hat{\boldsymbol{\nu}}_i = (\cos \theta_i, \sin \theta_i)$.  The
$\eta$ variables introduce Gaussian noise, with $\langle
\eta_i(t)\rangle = 0$ and $\langle \eta_i(t) \eta_j(t')\rangle =
\delta_{ij} \delta(t - t')$. Although the noise may be non-thermal, we set $\Dr = \frac{3D}{\sigma^2}$ (with $\sigma$ the particle diameter) as would apply to a sphere in the low-Reynolds-number regime.


Due to the self-replenishing velocity $\vp$, collisions between particles are rendered persistent, which leads to cluster formation from the dilute phase. To model such cluster formation in equilibrium systems, CNT assumes a free energy  of the form $G(n)=\Delta \mu V(n) + \gamma A(n)$, with $n$ the number of molecules in the droplet, $V(n)$ and $A(n)$ the droplet volume and surface area,  $\Delta \mu$ the chemical potential difference between the dense and dilute phases, and $\gamma$ the surface tension. This assumption, together with the Becker-D\"{o}ring description of cluster growth kinetics \cite{Becker1935}, allows predicting nucleation rates as a function of material constants and concentrations.

Since ABP phase separation is intrinsically nonequilibrium, the same prescription cannot be directly applied. Therefore, we take the opposite approach, starting from the Becker-D\"{o}ring kinetics and inferring an effective free energy landscape.
To this end, we assume that the state of an ABP system may be represented at the mesoscopic level by the number density $\rho_n$ of clusters with $n$ particles (previous studies \cite{Redner2013,Redner2013a,Richard2016} have shown that polarization of particle orientations on cluster peripheries is also an important reaction coordinate; this effect enters implicitly in our kinetic model below).
The evolution of $\rho_n$ is then given by a hierarchy of master equations accounting for events such as cluster growth, depletion, merging, and fragmenting. In our case, simulations additionally show that the system is well mixed and clusters evolve primarily through gain and loss of individual monomers from their perimeters. Under these conditions, the master equations take on the simple form:
\begin{align}
\partial_t \rho_n &= J(n-1) - J(n)
\label{eq:model_master} \\
J(n) &= \jin(n) S(n) \rho_n - \jout(n \negthinspace + \negthinspace 1) S(n \negthinspace + \negthinspace 1) \rho_{n+1}
\label{eq:model_flux}
\end{align}
where the fluxes $\jin$ and $\jout$ represent the rates of monomer gain and loss per unit of cluster surface and $S(n)$ is the surface area (perimeter) of a cluster of size $n$.

To gain insight into the phase behavior and kinetics of the system, we first consider a steady-state in which the fluxes $J$ are zero. While the existence of such steady-states in the physical system is not guaranteed, they are consistently observed in simulations \cite{Fily2012,Redner2013,Redner2013a,Stenhammar2013,Buttinoni2013a,Mognetti2013,Stenhammar2014b,Wysocki2013}. In the absence of phase separation, the steady-state corresponds to free monomers coexisting with small, transient clusters.  The steady-state cluster size distribution (CSD) $P(n)$ can be calculated by iterating Eq. \ref{eq:model_flux}:
\begin{equation}
  \rho_n = \rho_1 \prod_{m=1}^{n-1} \frac{\jin(m) S(m)}{\jout(m+1)
    S(m+1)} \equiv \rho_1 P(n)
  \label{eq:model_cluster_size},
\end{equation}

Within the phase separation regime, our simulations identify a parameter range within which nucleation is slow in comparison to the settling time of the CSD (analogous to the metastable region between the binodal and the spinodal in an equilibrium system). In these cases the CSD may still be taken to be (quasi) stationary, $\partial_t \rho_n= 0$, but to access the slow nucleation dynamics we now must acknowledge a small nonzero flux $J(n)$.
Under these conditions the fluxes $J$ are equal and given by \cite{SIref}:
%
\begin{align}
  J &= \rho_1 \left( \sum_{n=1}^\infty \frac{1}{\jin(n) S(n) P(n)} \right)^{-1}
  \label{eq:model_nucleation_rate}.
\end{align}
The mean nucleation time in a system with
volume $V$ is then $\taunucl = (J V)^{-1}$ \cite{Oxtoby1992}.

\begin{figure}
  \raisebox{17pt}{\includegraphics[width=.36\linewidth]{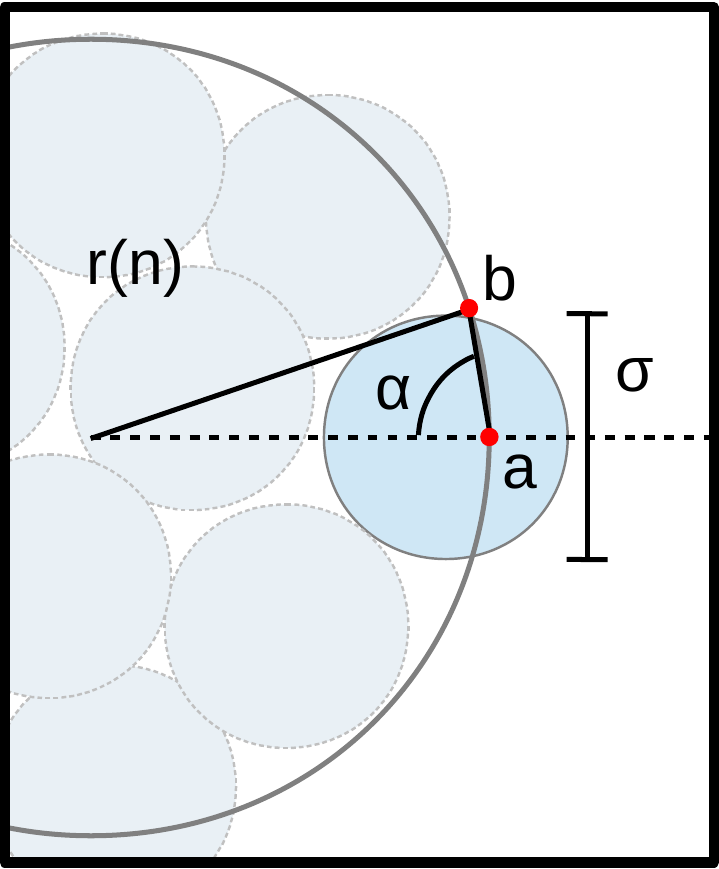}}
  \hfill
  \includegraphics[width=.61\linewidth]{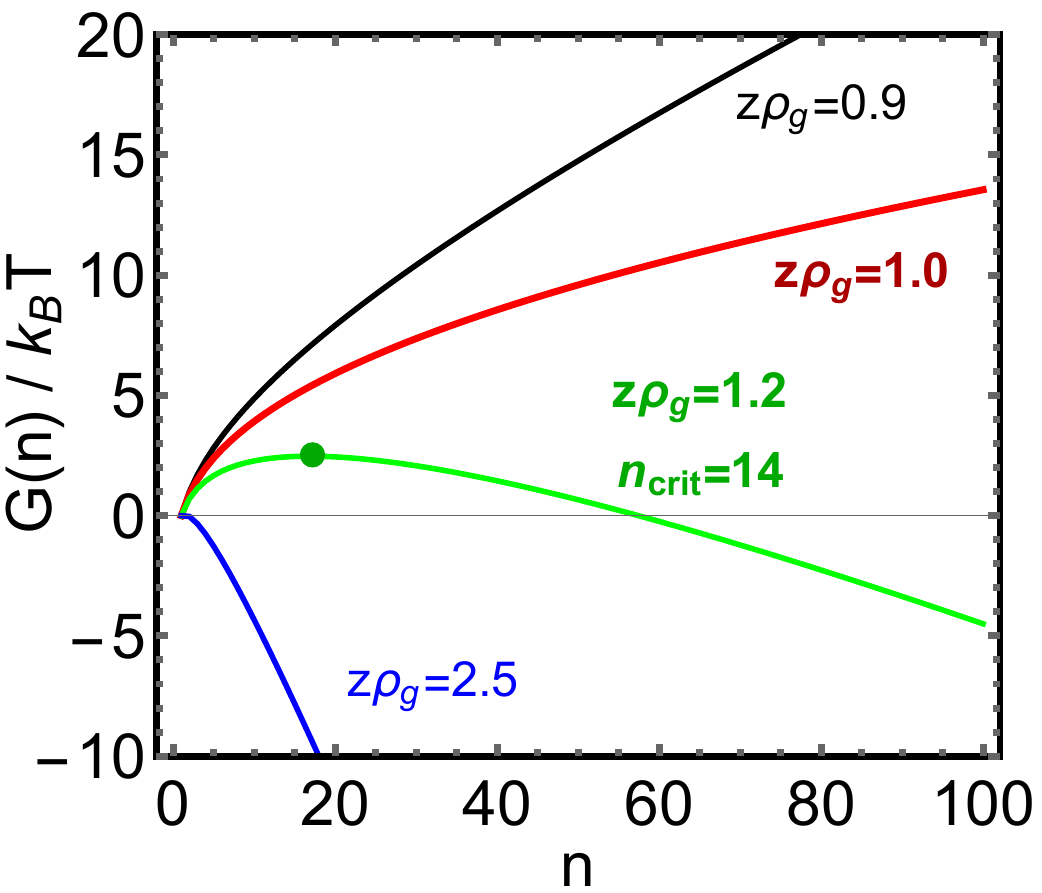}
  \caption{Left: Depiction of the influence of curvature. The angle $\alpha$ represents the width of the
    ``horizon'' above which a particle's propulsion must point in
    order for it to escape.  Right: Effective free energy as a
    function of the supersaturation $z \rhog$, in the single-phase
    region (black), at the binodal (red), in the metastable region showing a nucleation barrier (green), and in the spinodal
    regime (blue).}
  \label{fig:fig1}
\end{figure}

To proceed further, we construct a minimal microscopic
model that enables estimating the adsorption and evaporation fluxes.  We model each cluster as circular, with volume $V(n) =
n / \rhoc$ and surface area $S(n) = \sqrt{4 \pi n / \rhoc}$, with $\rhoc$ near the close-packing density for spheres. Using basic arguments for how particles adsorb on and depart from clusters \cite{Redner2013,Redner2013a,SIref}, we obtain $\jin =
\frac{\rhog \vp}{\pi}$ and $\jout(n) = \frac{\Dr}{\sigma} \left(\frac{\pi}{2 \alpha(n)} \right)^2$, where $\rhog$ is the monomer density in the ``free
volume'' not occupied by clusters (different from $\rho_1$), and $\alpha(n) = \frac{1}{2} \left( \pi -\sin^{-1} \frac{\sigma}{2 r(n)} \right)$ is the ``horizon angle" taking into account cluster curvature (Fig. \ref{fig:fig1}) \cite{Lee2015}.

Plugging in to Eq. \ref{eq:model_cluster_size}, we have:
\begin{align}
P(n) = \frac{(z \rhog)^{n-1}}{\sqrt{n}}  P_0(n)
\label{eq:Pn}
\end{align}
where $z = \frac{\vp \sigma}{\pi \Dr}$ 
is analogous to the P\'eclet
number (Pe), and $P_0(n) = \left[ \prod_{m=1}^{n-1} (2
\alpha(m+1)/\pi) \right]^2$ accounts for the geometric effects of cluster size.


At this stage the kinetic theory is complete, and we can use the above formulae directly to compute quantities of interest. On the other hand, we may continue the analogy with CNT by considering an effective free energy $G(n)$ (analogous to the grand potential in equilibrium statistical mechanics). Since in equilibrium fluctuation theory we would have $\rho_n \propto \exp\left[-G(n)/(k_B T) \right]$, we write
\begin{align}
G(n) = -k_B T \ln \left( \sigma^2 \rho_n \right).
\label{eq:model_free_energy}
\end{align}
Defined as such, $G(n)$ serves as a natural formulation of the kinetic theory in terms of a
functional landscape.
In particular, we note that the existence of such an effective free energy landscape depends only on the presence of a (quasi)-steady state (Eq. 5), and is not contingent on a thermodynamic interpretation.


\textit{Results.}  Working within the effective free energy picture, we use  Eqs. \ref{eq:Pn} and \ref{eq:model_free_energy} to obtain
\begin{equation}
  \Gex(n) = - \kt \negthinspace \left[ n \ln (z \rhog) - \frac{1}{2} \ln (n) + \ln \left(P_0(n) \right) \right]
  \label{eq:kinetic_free_energy}
\end{equation}
where terms not depending on $n$ have been dropped. It is evident that the quantity $z \rhog$ controls the phase
behavior of the system, and is analogous to the supersaturation of the
fluid phase.  As shown in Fig. \ref{fig:fig1},
$\Gex(n)$ is monotonically increasing when $z \rhog < 1$,
corresponding to a homogeneous fluid, while for $1 < z \rhog <
\sqrt{2} \left( \frac{2 \alpha(2)}{\pi} \right)^2 \approx 2.42$ it
exhibits a barrier followed by a monotonic decrease, corresponding to
a supersaturated fluid which is metastable to cluster nucleation. At higher values (beyond the `spinodal'), $\Gex(n)$ is monotonically
decreasing and the system is unstable towards cluster formation.

For large clusters, $P_0$ can be simplified to (see section D of \cite{SIref}):
%
%
\begin{equation}
\Gex(n) = - \kt \left[ \ln(z \rhog) n  - \frac{\sigma \rhoc}{\pi} S(n) \right] + \mathcal{O}(\ln n)
\label{eq:kinetic_free_energy_2}
\end{equation}
%
From this,
we see that a geometric understanding of ABP
microdynamics leads naturally to their equilibrium-like phase behavior.
Based on its role in governing the phase behavior of the system, the first term in Eq. \ref{eq:kinetic_free_energy_2} controls the relative propensity of a particle to be in the dilute or dense phase. Thus, in analogy with the equilibrium CNT free energy, this term represents
the difference in effective chemical potential $\Delta \mu$ between the two
phases.
Our expression $\Delta \mu = \ln( \frac{\vp \sigma}{\pi \Dr} \rhog)$ has similar structure to the chemical potential $\Delta \mu = \ln(\rho) + \ln\left[\vp(\rho)\right]$ ($\rho$ being a coarse-grained density field) considered in the continuum theory of Stenhammar, et. al \cite{Stenhammar2013}. The crucial difference here is the explicit appearance of the microscopic diffusion constant $\Dr$ in place of the functional ansatz $\vp(\rho)$. 
The second term in Eq.~\ref{eq:kinetic_free_energy_2} is related to the cluster's surface area, and can be interpreted as an effective line tension that drives coarsening.  Note that in a nonequilibrium system, this need not equal the mechanical line tension, and in fact Bialke et al.~\cite{Bialke2015} measured a negative mechanical line tension for a flat interface in an ABP system. Because the excess free energy associated with interface formation must be positive for stability, Bialke et al.~suggest the negative line tension is balanced by a positive interfacial stiffness.  Since the line tension emerges from our calculation as a consequence of cluster curvature (expressed through the horizon angle $\alpha(n)$) there may be a connection to the Bialke et al. measurement, but comparisons at additional values of Pe are needed to explore this possibility.
Finally, solving $\Delta G(n) =
0$ gives a prediction for the critical
nucleus size as $\ncrit = \sigma^2 \rhoc / \pi \left[\ln(z \rhog)
  \right]^2$, a form familiar from CNT.


\begin{figure}
  \includegraphics[width=.75\linewidth]{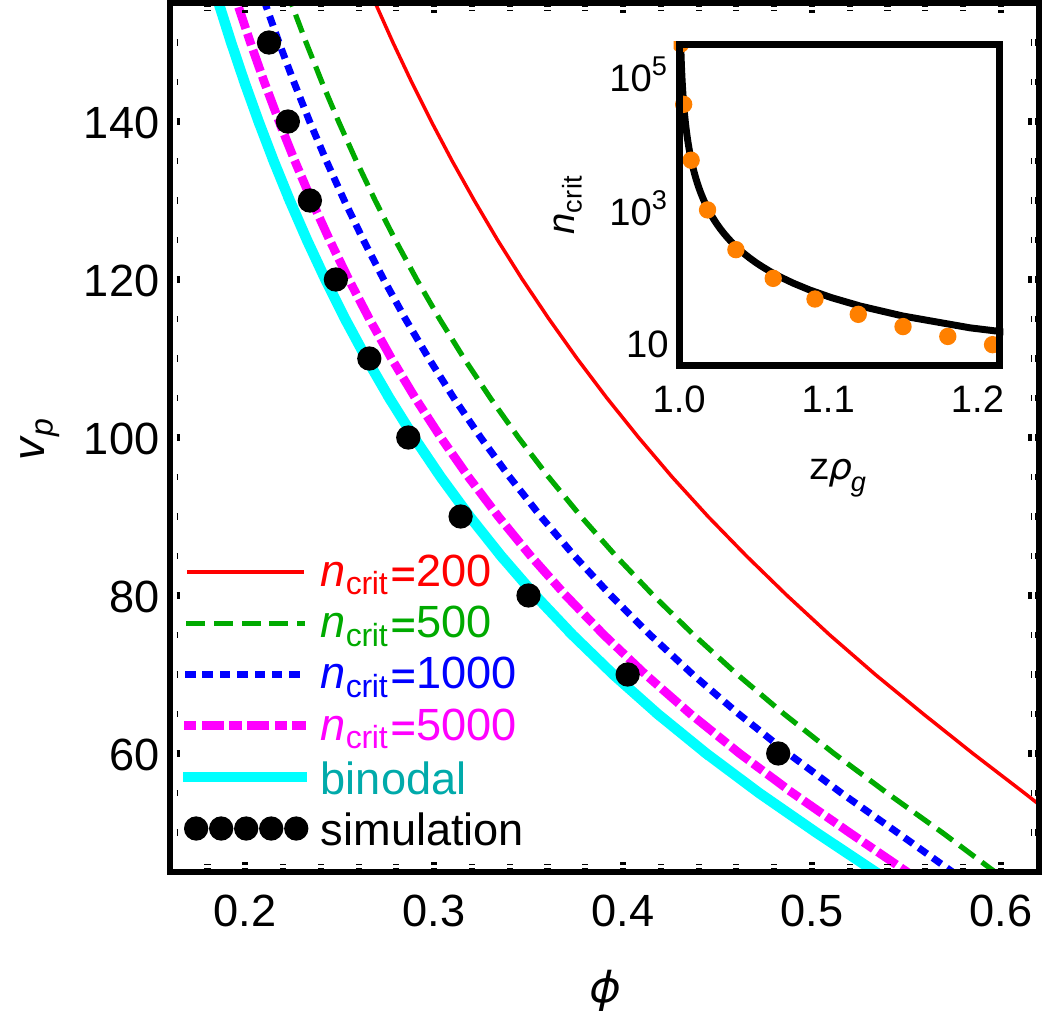}
  \caption{Plot of the lower binodal and a few iso-critical lines as
    computed from our kinetic theory.  Black dots denote the location
    of the binodal as measured from simulations, showing remarkable
    agreement with the theoretical prediction.
 \textit{Inset:} An expanded view of the phase diagram, showing additionally the upper binodal (black dots) and the lower spinodal (red squares) as measured  from simulations \cite{SIref}. The open circles denote systems observed in simulations to be single-phase, thereby demarcating the approximate location of the phase boundary near the critical point (dashed black line).  }
  \label{fig:phase-diagram}
\end{figure}


Next, we demonstrate the quantitative insights of the theory into ABP systems.
Our simulations are performed as in \cite{Redner2013} (see \cite{SIref} for details), though to measure nucleation times much larger than those accessible in brute force simulations,
 we used a weighted-ensemble dynamics \cite{Huber1996,Zhang2010,Zhang2007}. To limit finite-sized effects in the NVT simulations, we consider systems with $15000$ particles, thus providing a good estimate of nucleation rates except very near the binodal where the critical nucleus size approaches the system size.  \footnote{We use NVT simulations rather than NPT (as would be typical in an equilibrium investigation of CNT) because the relevant experiments correspond to the NVT ensemble (\eg \cite{Palacci2013a}).  Furthermore, defining the NPT ensemble requires a thermodynamic pressure. While such a pressure can be defined for the system under consideration \cite{Solon2015,Solon2015b,Takatori2014,marchetti2015structure,Takatori2015,Speck2015,Speck2015a,Winkler2015,Ginot2015}, its use to implement NPT simulations remains as of yet untested.  Our system sizes are large enough to avoid noticeable deviations from the thermodynamic limit which can occur in NVT simulations with extremely small system sizes \cite{Mayer1965,Thompson1984,Alder1962,Reguera2003}.} Also, since the control parameter in our NVT simulations is the overall volume fraction $\phi$ whereas in the theory it is the theoretical  gas density $\rhog$, we must construct a coordinate transformation which relates the two \cite{SIref}:
\begin{align}
\phi = \frac{A \rhoc}{\left(4/\pi \sigma^2\right)A- 1 + \rhoc/\rho_{g}}
\label{eq:transformation}
\end{align}
where $A=\frac{\pi \sigma^{2}}{4}\sum_{n=1}^{\nmax} n P(n)$ with $\nmax$ a cutoff cluster size \cite{SIref}. Finally, to make our comparison with simulations quantitative, we empirically fit the upper binodal by measuring the density within large clusters, $\rhoc(\vp)$, which is found to increase with $\vp$ due to the imperfectly hard interaction potential \cite{SIref}. The resulting phase diagram is shown in Fig. \ref{fig:phase-diagram}. The predicted lower binodal is remarkably close to its measured location, although this could be partly fortuitous.
%

\begin{figure}
  \includegraphics[width=.75\columnwidth]{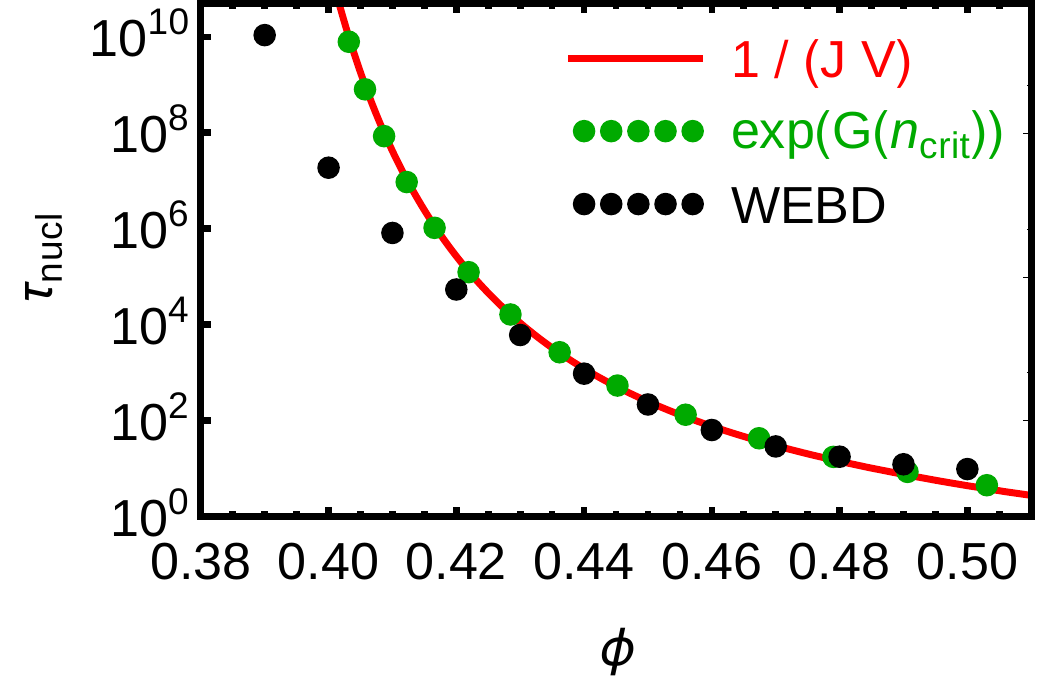}
  \caption{Mean nucleation times as computed from the full kinetic
    expression (Eq. \ref{eq:model_nucleation_rate}, red), and by
    applying an Arrhenius form to the height of the nucleation barrier
    (Eq. \ref{eq:kinetic_free_energy}, green).  The latter approach
    does not supply the kinetic prefactor, which must be determined by
    fitting. Nucleation rates as measured by weighted ensemble dynamics
    simulations are shown in black.}
  \label{fig:nucleation}
\end{figure}
We now employ Eqs. \ref{eq:transformation} and
\ref{eq:model_nucleation_rate} to compute the mean nucleation
time in the metastable regime.  Here we find that the theory, thus far constructed without adjustable parameters, lacks quantitative
accuracy because the predicted nucleation rate is
exquisitely sensitive to small perturbations of $z \rhog$. To enable
qualitative comparisons, we set $\rhog^\text{eff} = \kappaNew \rhog$,
with $\kappaNew$ a fitting parameter that adjusts the monomer
chemical potential.  We find (by eye) that setting $\kappaNew=0.71$
produces good correspondence with simulation (Fig. \ref{fig:nucleation}).   Nucleation rates are notoriously difficult to
quantitatively predict from first principles even in the equilibrium
case (\eg \cite{Auer2001,Prestipino2012,Loeffler2013,Statt2015,Mcgraw1997}), so the correspondence between
theory and simulation with a single small fitting parameter is
notable.

\begin{figure}
  \includegraphics[width=.49\linewidth, height=.49\linewidth]{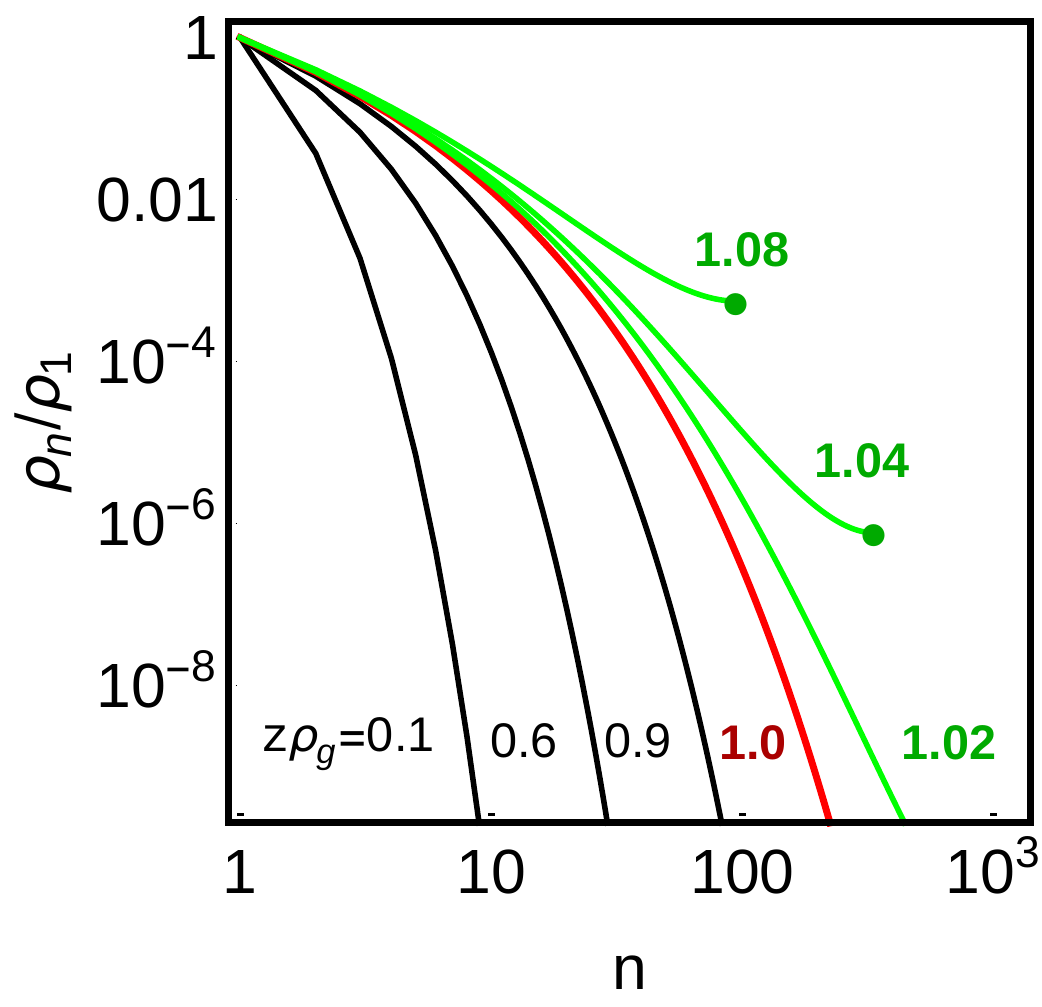}
  \includegraphics[width=.49\linewidth, height=.49\linewidth]{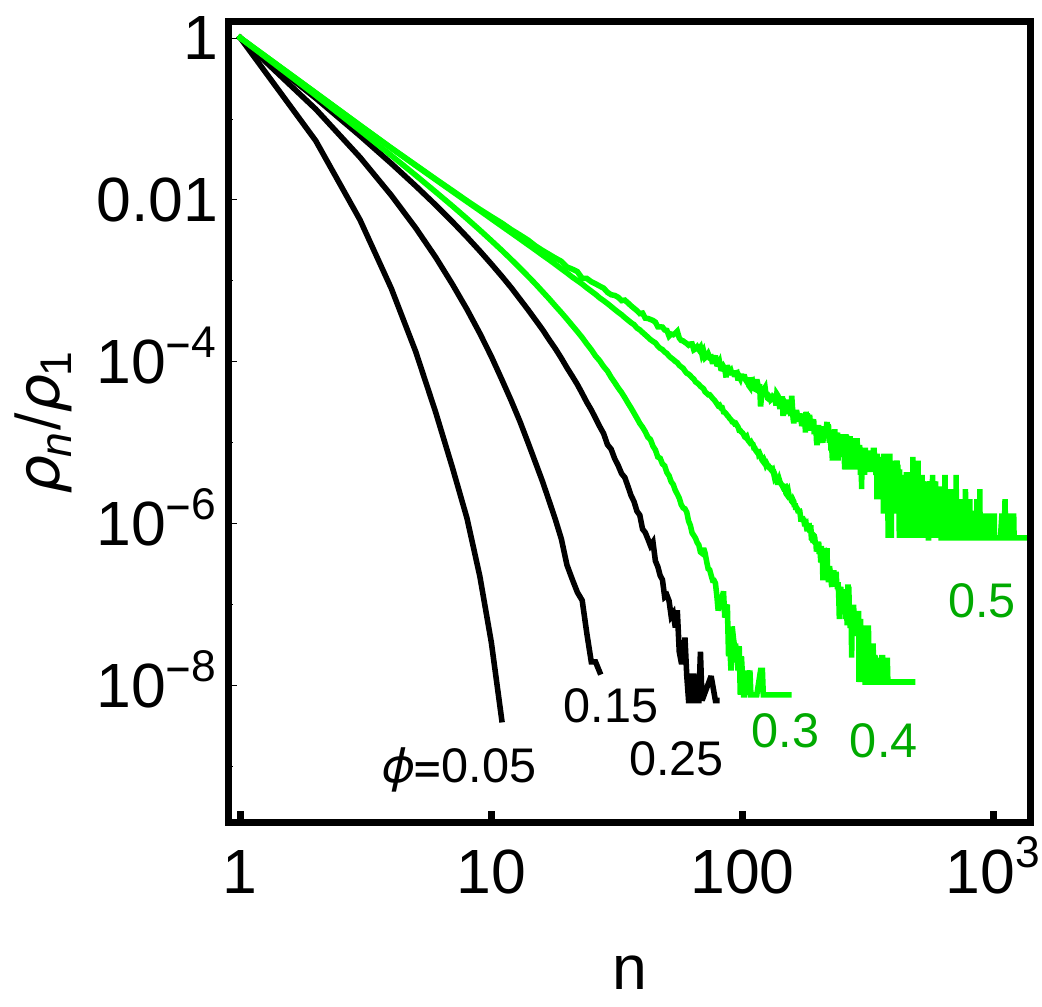}
\includegraphics[width=.49\linewidth, height=.49\linewidth]{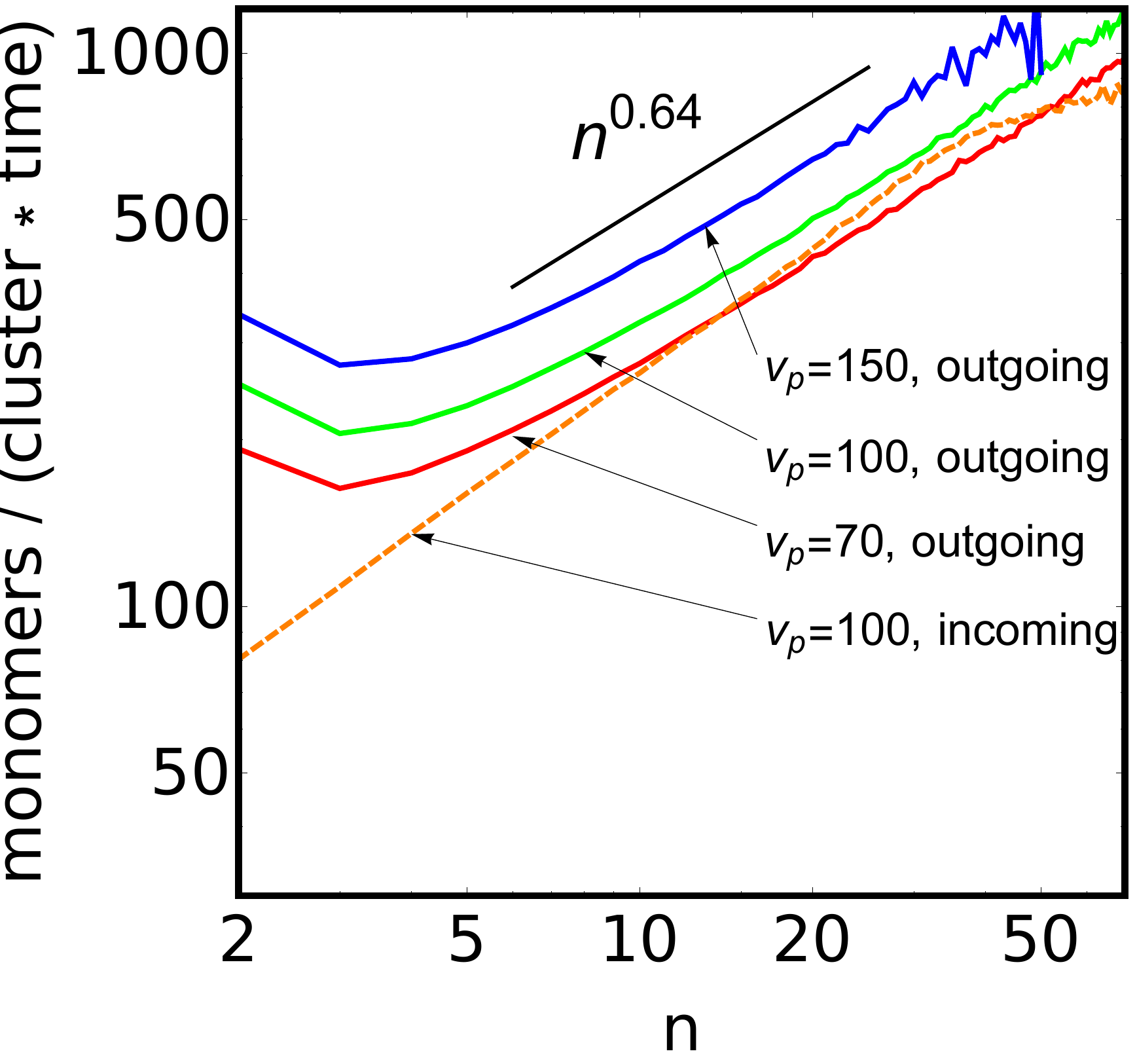}
\includegraphics[width=.49\linewidth, height=.49\linewidth]{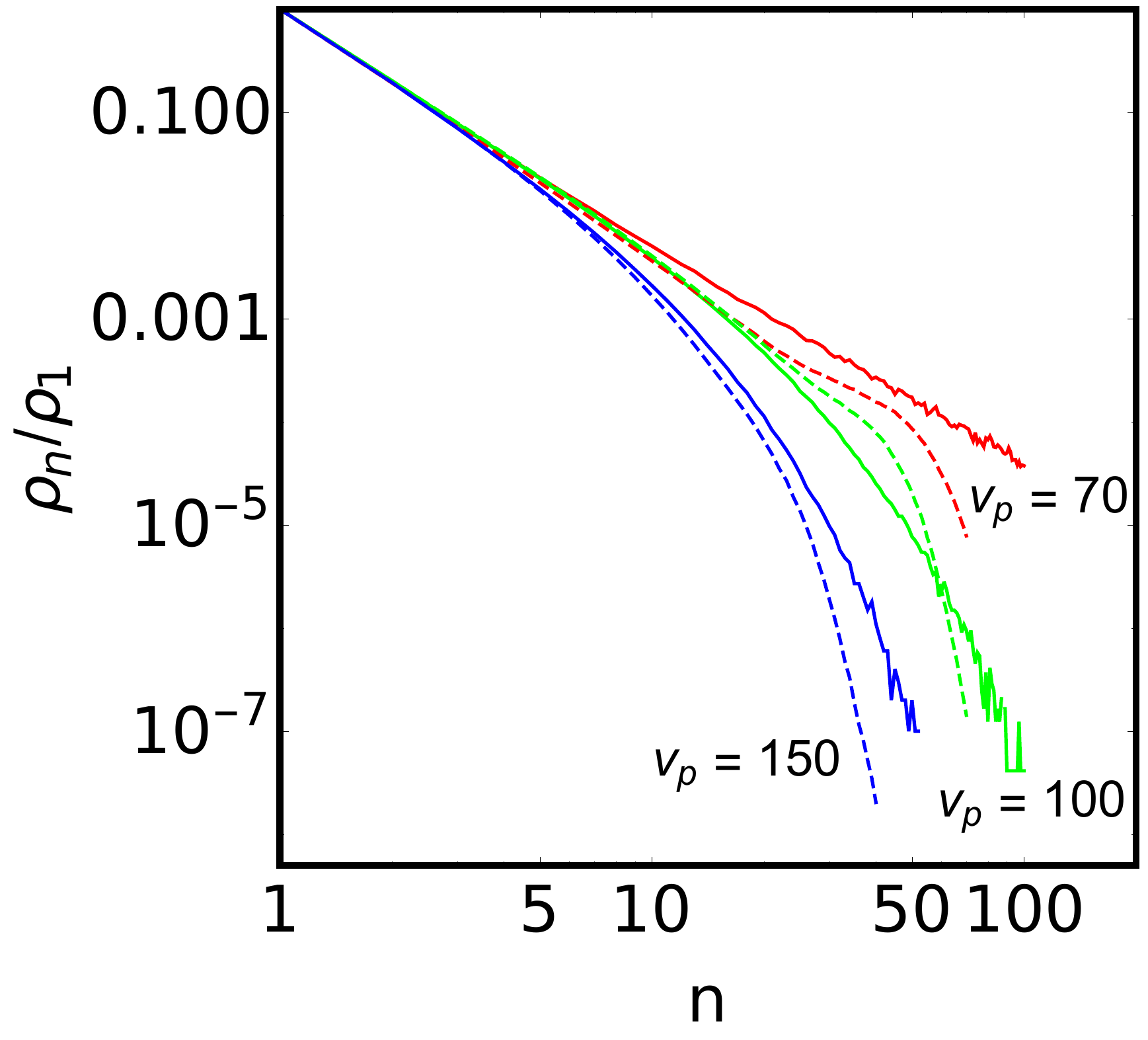}
  \caption{
 Top: Cluster size distributions (CSDs), as predicted by Eq. \ref{eq:model_cluster_size} (left) and as measured in simulations for $\vp = 100$ (right). Data are colored based on their location in the phase diagram: black for single-phase, red at the binodal, and green in the metastable regime. Bottom left: Averaged incoming and outgoing rates in units of (monomers/(cluster*time)) from simulations above the binodal, with $z \rhog = 1.04$. Bottom right: Comparison of the CSDs at $z \rhog = 1.04$ from simulations (solid) and as reconstructed from the simulated rates (dashed).}
  \label{fig:rhon}
\end{figure}
Finally, we compare the theoretical and simulated CSDs in Fig.~\ref{fig:rhon}. Below and slightly above the binodal, the theory has the correct functional form, but far into the metastable region fails to account for power law scaling $\rho_n \propto n^{-2}$  below a threshold cluster size, indicative of logarithmic corrections to $G(n)$ \footnote{While the theoretical CSD does predict a power law (represented by the logarithmic correction to $G(n)$ in Eq. \ref{eq:kinetic_free_energy_2}), it is suppressed in the region of interest and has exponent ${\raise.17ex\hbox{$\scriptstyle\mathtt{\sim}$}} 0.59$ instead of $2$.}. Similarly,   Levis et al.~\cite{Levis2014}, observed power law scaling with exponent $1.70 \pm 0.05$  in a related model system.

In equilibrium CNT, logarithmic corrections in the free energy have been obtained by accounting for degrees of freedom internal to a cluster, representing deviations from a spherical shape (\eg \cite{Reguera2001,Kiang1971,Fisher1967}). Our results suggest analogous mechanisms play a role here. Simulations show that clusters have ramified structure, with a fractal surface area scaling $S(n) {\raise.17ex\hbox{$\scriptstyle\mathtt{\sim}$}} n^{0.64 \pm 0.01}$ \cite{SIref}. Indeed, directly measuring fluxes (Fig.~\ref{fig:rhon}) shows deviations from what is expected for large spherical clusters: Simulated rates are larger than predicted by the theory, with the outgoing rates depending on $\vp$. These attributes are consistent with complex cluster geometry, since on small clusters or near regions of high curvature, particles can escape by ``sliding off" each other before completely rotating to the horizon, thus resulting in a higher than predicted, $\vp$-dependent outgoing rate. Multi-particle escape events may also contribute \cite{Redner2013}. 
To test whether these mechanisms are responsible for the power law scaling, we fed the measured rates into Eq.~\ref{eq:model_cluster_size} to reconstruct the CSDs, which showed good agreement with simulation (Fig.~\ref{fig:rhon}). Similar results were obtained elsewhere in the phase diagram (see \cite{SIref}).  Thus, a calculation of the rates which takes these effects into account should recover the scaling of simulated CSDs.

In summary, we have shown that an approach analogous to classical nucleation theory can describe the nonequilibrium nucleation of clusters from solutions of ABPs. By linking the microscopic parameters of ABPs to their macroscale kinetics, this framework makes an important step toward developing design principles for applications of motility-induced phase separation \footnote{While completing this manuscript, we learned of another concurrent study which computationally measures ABP nucleation rates \cite{Richard2016}. By analyzing simulation trajectories, they conclude that cluster size and cluster polarization (the extent to which particles on the rim of a cluster point inwards) are important collective coordinates for describing transition rates. These observations are consistent with the kinetic theory described here.}.

\FloatBarrier

{\bf Acknowledgments.} This work was supported by the NSF (DMR-1149266 and the Brandeis MRSEC DMR-1420382).
Computational resources were provided by the NSF through XSEDE computing resources and the
Brandeis HPCC.



\newpage

\section{A classical nucleation theory description of active colloid assembly \\
  Supplementary Materials}

\affiliation{Martin Fisher School of Physics, Brandeis University,
  Waltham, MA, USA.}

\thispagestyle{fancy}
\pagestyle{fancy}

\subsection{Simulation Details}
{\bf Cluster size distributions.} To validate the predicted cluster size distributions (CSDs) and phase diagram,
we performed Brownian dynamics simulations identical to those in
\cite{Redner2013}.  To measure the CSDs, we
performed simulations with $N=30000$ particles, run for $200 \tau$.
Particles were considered to be bonded if their separation was less than
$\sigma$, and clusters were identified as groups of bonded particles
\cite{Redner2013, Redner2013a}.  We ran five simulations at each
parameter set, discarded the first $1 \tau$ of each trajectory, and
averaged the results.  We also discarded frames in which any cluster
containing more than $1500$ particles existed, to exclude
phase-separated configurations.

\begin{figure}
 \includegraphics[width=.85\linewidth, height=0.75\linewidth]{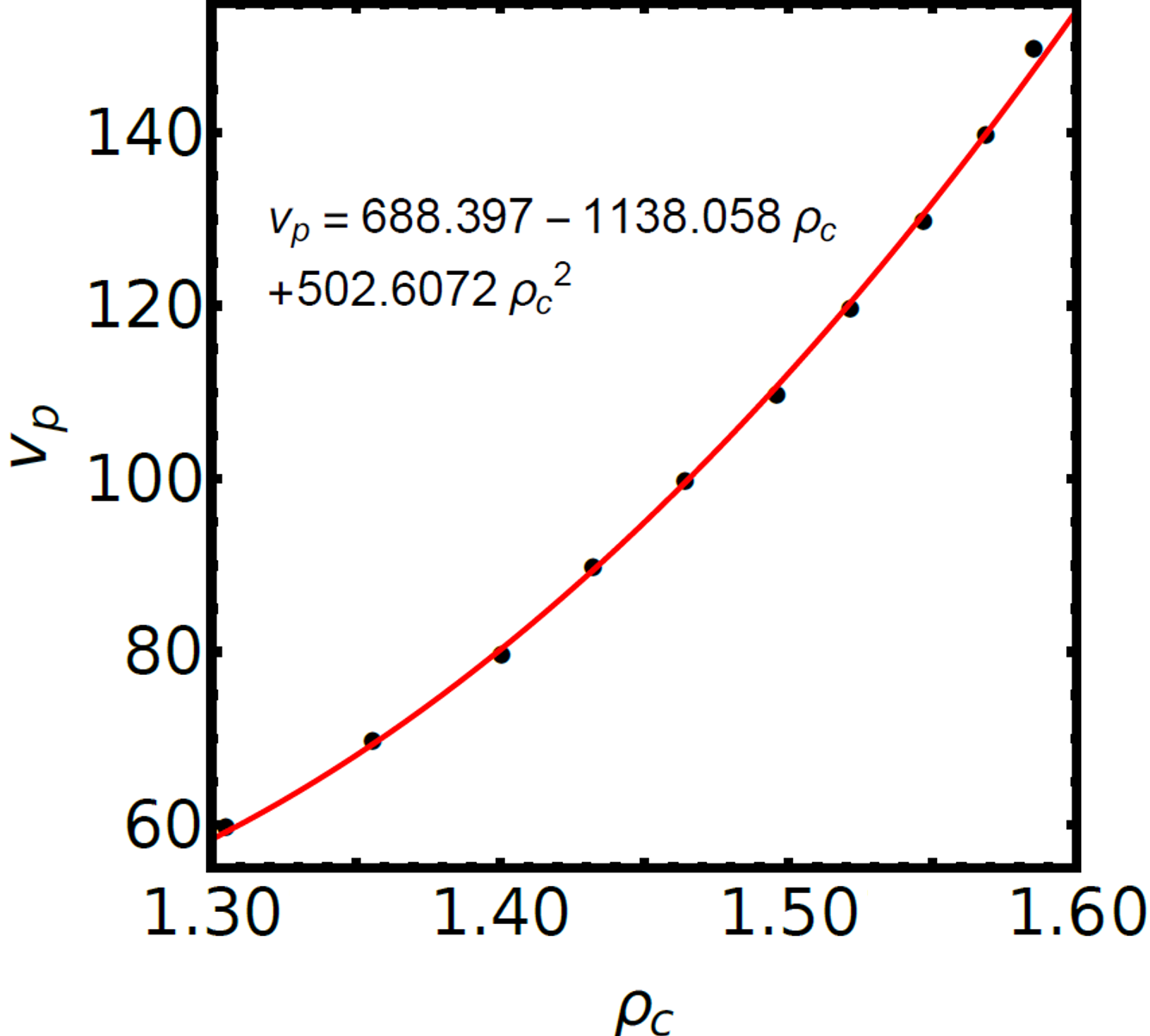}
 \caption{Cluster densities $\rhoc$ as a function of Peclet number for phase-separated systems. Inversion of the quadratic fit (red) gives Eq.  \ref{eq:rhoc_fit}.}
  \label{fig:rhoc_fit}
\end{figure}

{\bf Binodals.} To locate the upper and lower binodal, we ran similar simulations until $400 \tau$ and
measured the density of the cluster interior $\rho_c$ (discarding the edge
region) and the density of the dilute phase (discarding the cluster
and surrounding boundary region). Each data point was averaged over ten separate
simulations. The results for the upper binodal are shown in Fig. \ref{fig:rhoc_fit}, with the resulting trend fit to the functional form
\begin{align}
\rho_c \approx 1.1322 + \sqrt{0.00199 \vp - 0.08788}.
\label{eq:rhoc_fit}
\end{align}

\begin{figure}
 \includegraphics[width=1.0\linewidth, height=0.75\linewidth]{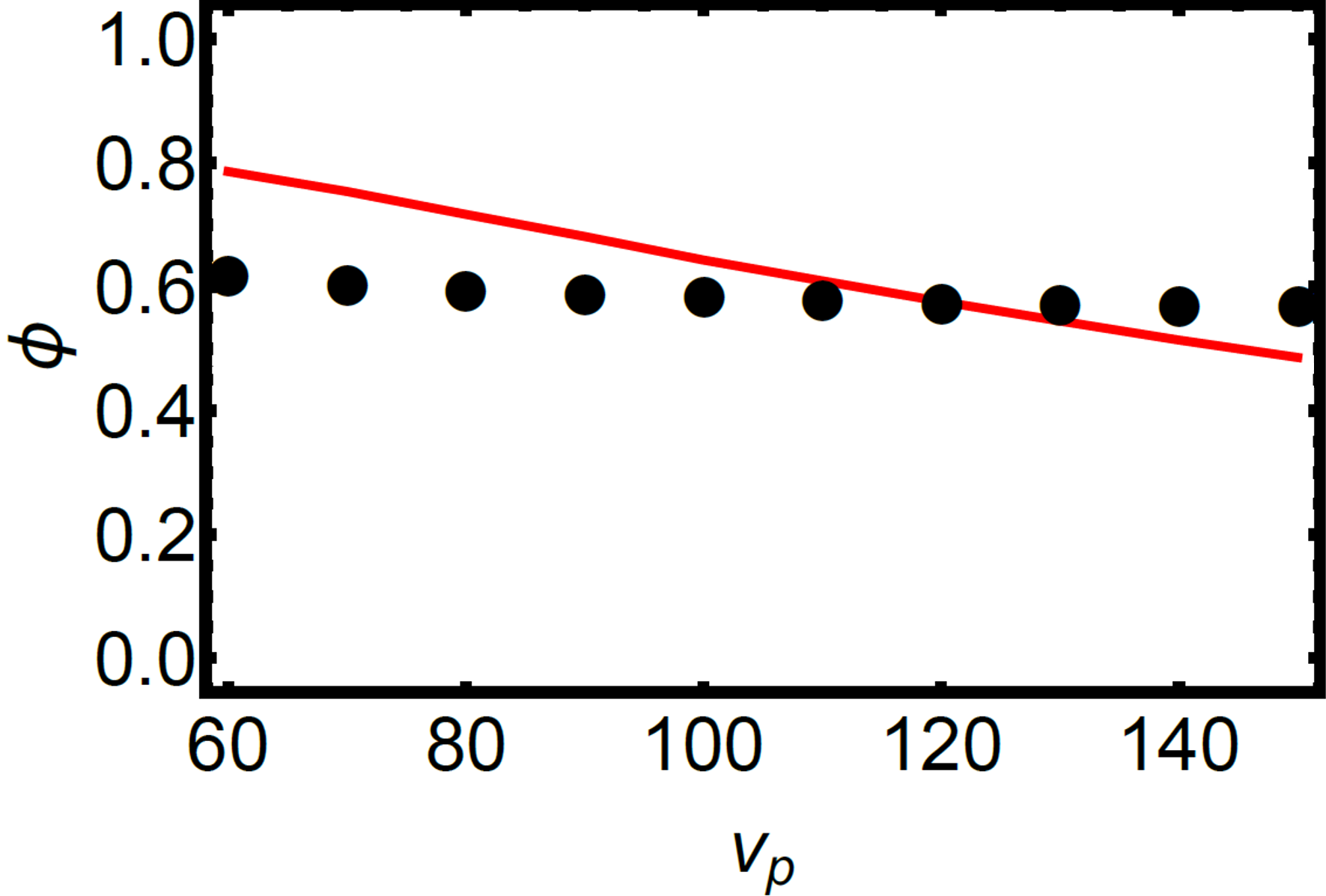}
 \caption{The approximate location of the lower spinodal measured in simulations (black dots) and estimated by the theory (red) as described in section A. The discrepancy between the theoretical and simulation estimates can be explained by noting that assumptions in the theory, such as neglecting interactions between clusters and the occurrence of a quasi-stationary CSD, break down near the spinodal (see section A).}
  \label{fig:lower_spinodal}
\end{figure}

\def\ns{n_\text{s}}
\def\taus{\tau_\text{s}}
\textbf{Spinodals.} To estimate the lower spinodal from simulation trajectories, we set the criterion that a system produce a large cluster (with $n \ge n_\text{s}$ subunits) within a small threshold time $\tau_s$. The spinodal is then approximated by the locus of points in phase space which produce, on average, a large cluster after exactly $\tau_s$ time units. For these measurements, we set $\ns=1000$ and $\tau_s = 0.1 \tau$. Since each particle travels on the order of 10 particle diameters in this time interval, we expect it to reasonably estimate the time required for a large cluster to assemble in a system that is unstable towards cluster formation. Moreover, visual inspection confirms that this criterion roughly marks the transition in phase space from the nucleation of isolated clusters to spinodal decomposition. The results of this procedure are shown in Fig. \ref{fig:lower_spinodal}, with each data point averaged over 50 simulations.

We also attempt to estimate the spinodal from the theory. In this context, the spinodal corresponds to the locus of points in phase space for which $n_{crit} = 0$, which in our case occurs at $z \rho_g \approx 2.42$, as shown in the discussion following Eq. \ref{eq:kinetic_free_energy}. To turn this into a prediction in the ($v_p$, $\phi$) phase diagram, we must use the coordinate transform between $\rho_g$ and $\phi$ (Eq. \ref{eq:transformation}). However, this relation assumes a stationary CSD as given by Eq. \ref{eq:Pn}, an assumption which fails as $n_{crit}$ goes to $0$. Indeed, for small values of $n_{crit}$ ($\lesssim 20$), the coordinate transform changes its behavior, such that decreasing density leads to decreasing $\ncrit$, which is clearly unphysical. To nevertheless obtain a rough estimate of the spinodal in terms of $\phi$, we have altered the criterion for the spinodal from $\ncrit = 0$ to $\ncrit \approx 30$, in which case the coordinate transform is still usable. As expected, this prediction does not closely match the measured spinodal (Fig. \ref{fig:lower_spinodal}). More broadly, we expect any theory which relies upon a separation of timescales between nucleation and cluster growth, or which does not explicitly account for interactions between clusters, to break down near the spinodal. In this regime the dense phase forms through the simultaneous nucleation and merging of many small clusters.

\textbf{Measurement of nucleation rates in simulations.}
The rates of nucleation were determined using the weighted-ensemble
(WEBD) method \cite{Huber1996,Zhang2010,Zhang2007}, which allows for precise rate
measurement of extremely rare events and is applicable to non-equilibrium dynamics.  We performed the calculation on systems of size
$N=15000$ and used the size of the largest cluster as our progress coordinate.
At the first appearance of a cluster of size $n_{0}$, a system was considered to have undergone a nucleation event. We set $n_0=3000$; its precise value does not affect the results provided it is larger than the critical nucleus size. We ran each ensemble for 160 to 2000 WEBD
iterations, depending on the rate of convergence. The bin boundaries were calculated separately for each WEBD run using
the procedure outlined in Ref.~\cite{Huber1996}, Appendix C

\subsection{Calculation of Nucleation Rate}

We assume the cluster size distribution remains (quasi) stationary, such that $\partial_t
\rho_n = 0$ and therefore $J(n) = J$, where $J$ is a constant. Using the definition of $J(n)$, this implies
\begin{align}
\rho_{n+1} - b_{n}\rho_{n} = - \frac{J}{ \jout(n \negthinspace + \negthinspace 1) S(n \negthinspace + \negthinspace 1)}.
\end{align}
where $b_{n}\equiv \frac{\jin(n) S(n)}{\jout(n \negthinspace + \negthinspace 1) S(n \negthinspace + \negthinspace 1)}$. Now dividing both sides by $\prod_{m=1}^{n} b_{m}$, we get:
\begin{align}
\frac{\rho_{n+1}}{\prod_{m=1}^{n} b_{m}} - \frac{\rho_{n}}{\prod_{m=1}^{n - 1} b_{m}} = - \frac{J}{\jin(n) S(n) \prod_{m=1}^{n - 1} b_{m}}.
\end{align}
Summing from $n = 1$ to an arbitrary cutoff $n = N - 1$ gives
\begin{align}
\frac{\rho_{N}}{\prod_{m=1}^{N - 1} b_{m}} - \rho_{1} = -J \sum_{n=1}^{N - 1} \frac{1}{\jin(n) S(n) \prod_{m=1}^{n - 1} b_{m}}.
\end{align}
As $N \rightarrow \infty$, $\rho_{N} \rightarrow 0$. Taking this limit, and also recognizing that $\prod_{m=1}^{n - 1} b_{m} = P(n)$, we obtain
\begin{align}
\rho_{1} = J \sum_{n=1}^{\infty} \frac{1}{\jin(n) S(n) P(n)}.
\end{align}
And so
\begin{align}
J = \rho_{1} \left( \sum_{n=1}^\infty \frac{1}{\jin(n) S(n) P(n)} \right)^{-1}.
\end{align}

\subsection{Kinetic Theory}

As mentioned in the text, to estimate $\jin$ and $\jout$ we first model each cluster as circular, with volume $V(n) =
n / \rhoc$ and surface area $S(n) = \sqrt{4 \pi n / \rhoc}$, comprised
of nearly close-packed spheres at number density $\rhoc$.

To estimate the arrival flux $\jin$, we assume the monomer gas to be
homogeneous, isotropic, and composed of particles moving at a single
constant speed $\vp$.  We additionally assume $\vp \sigma / D \gg 1$ so
we may neglect the effects of translational diffusion. The flux
of monomers impinging on a cluster surface is then $\jin =
\frac{\rhog \vp}{\pi}$, independent of cluster size.

For the departure flux, we consider a stationary particle on
a cluster surface, with its orientation evolving diffusively.
The particle remains on the surface only so long as its propulsion is
oriented inwards to the surface, and so we solve the rotational
diffusion equation with absorbing boundaries at the ``horizon'' where the
propulsion is oriented outwards and the particle escapes. A particle probes the cluster surface over the
scale of its own diameter (see Fig. 1 in the main text), and we therefore
estimate the horizon angle as $\alpha(n) = \frac{1}{2} \left( \pi -
\arcsin \frac{\sigma}{2 r(n)} \right)$.  Following the approach of
\cite{Redner2013a}, we find
$\jout(n) = \frac{\Dr}{\sigma} \left(\frac{\pi}{2 \alpha(n)} \right)^2$.

(We note that a similar ``horizon effect" for the kinetics of active particles on a curved surface has been considered by Lee \cite{Lee2015}.)

{ \bf Calculation of $\jout$ for non-spherical clusters.}
While not playing a central role in our theory, leading order changes to $\jout$ due to cluster asphericity have a simple relation to the surface area scaling $\langle S(n)\rangle$. Since aspherical clusters can adopt a wide range of cluster morphologies, and in principle each configuration will result in a different $\jout$, the quantity of interest becomes an ensemble-averaged outgoing flux, $\joutbar(n)$:
\begin{align}
\joutbar(n) \equiv \frac{1}{\langle S(n)\rangle}\left\langle \oint_{S_{c}} \jout(n,s)ds \right\rangle,
\label{eq:nonspherc1}
\end{align}
where $\jout(n, s)$ is the local outgoing flux on a cluster surface $S_c$ parametrized by $s$, and angled brackets denote an average over the ensemble of clusters of size $n$.
For an arbitrary point on $S_{c}$ (excluding the degenerate case of 0 curvature) there is a unique osculating
circle whose radius $r(n,s)$ defines the curvature $k(n,s)$ at that point:
\begin{align}
r(n,s)=\frac{1}{\left\vert k(n,s)\right\vert }.
\end{align}
The absolute value can be removed provided we define a signed ``radius" $\rsgn(n,s)$ that matches the sign of the curvature:
\begin{equation}
\rsgn(n,s)=\frac{1}{k(n,s)}.
\end{equation}
The same calculation employed above to calculate the outgoing rate on a spherical cluster applies \textit{%
locally} on $S_c$, provided we take for our ``radius"\ the quantity $\rsgn(n,s)$. Using the above result for $\jout$, this allows us to write%
\begin{align}
& \oint_{S_{c}} \jout(n,s)ds =  \\
 & \frac{\pi ^{2}}{4}\frac{\Dr}{\sigma }\oint_{S_{c}} \left[
\frac{1}{2}\left( \pi -\arcsin \left( \frac{\sigma }{2\rsgn(n,s)}\right)
\right) \right] ^{-2}ds. \nonumber
\end{align}
To proceed further, we make the following approximation: Clusters with
regions of high curvature (such that $\rsgn(n,s)$ is smaller than or on
the order of $\sigma $) are improbable. With this assumption we can expand the
integrand to obtain
\begin{align}
\oint_{S_{c}} \jout(n,s)ds \simeq \frac{\pi ^{2}}{4}\frac{\Dr}{\sigma }\left[ \frac{4}{\pi ^{2}}S(n)+\frac{4\sigma }{\pi ^{3}}\oint_{S_{c}} k(n,s)ds\right].
\label{approx_jout}
\end{align}
The second integral may be evaluated using the Gauss-Bonnet theorem: for the topology of a disk, $\oint k(n,s)ds = 2 \pi$. Therefore,
\begin{align}
\oint_{S_{c}} \jout(n,s)ds = \frac{\Dr}{\sigma } S(n)+2\Dr.
\end{align}
Referring back to Eq. \ref{eq:nonspherc1}, we then obtain
\begin{align}
\overline{\jout}(n) = \frac{\Dr}{\sigma } +\frac{2\Dr}{\langle S(n)\rangle}.
\end{align}

Thus, the outgoing rate per cluster $\overline{\jout}(n) \langle S(n)\rangle$ should scale as $n^{0.64}$, which agrees reasonably well with the results shown in Fig. \ref{fig:rates}. We note, however, that this corrected expression for $\jout$ has only a small quantitative effect on the theoretical CSDs, and in particular is not sufficient to capture the power law phenomenon observed in simulated CSDs. Thus, it seems that the appearance of power law scaling is connected primarily to small cluster deviations or the effects of high curvature, both of which have been neglected in the approximation leading to Eq. \ref{approx_jout}.  For these reasons, the result here is not pursued further in the development of our model.

\subsection{Large-$n$ scaling of the free energy}
We wish to find the large-$n$ scaling of the quantity $\ln (P_0(n)) = 2 \sum_{m=1}^{n-1} \ln \left[ (2\alpha(m+1)/\pi) \right]$, where $\alpha(n) = \frac{1}{2} \left( \pi -\arcsin \frac{\sigma}{2 r(n)} \right)$ is the horizon angle defined in the main text.
Based on our model of clusters as circular structures with density $\rhoc$, $r(n) = \sqrt{n/\pi \rhoc}$, so
\begin{equation}
\ln (P_0(n)) = 2 \sum_{m=1}^{n-1} \ln \left[ 1 - \frac{1}{\pi} \arcsin \left( \sqrt{ \frac{\rhoc \sigma^2 \pi}{4(m+1)}} \right) \right]
\end{equation}
The scaling of this quantity for large $n$ may be studied by expanding the arcsine and approximating the sum with an integral (the latter step is justified because the integrand is slowly varying for large $m$):
\begin{equation}
\ln(P_0) \approx 2 \int_1^{n-1} dm \, \ln \left( 1 - x/\sqrt{m+1} \right)
\end{equation}
where $x = \sqrt{\rhoc \sigma^2 / 4   \pi}$. Evaluating the integral and keeping terms up to $\mathcal{O}(\ln n)$, we obtain
\begin{equation}
\ln(P_0) \approx -4 x\sqrt {n}-{x}^2\ln \left( n \right),
\end{equation}
which gives for the free energy
\begin{align}
&G(n) \approx  \\
&-\kt \left[ \ln(z \rhog) n  - \frac{\sigma \rhoc}{\pi} S(n) - \left(\frac{1}{2} +x^2 \right) \ln(n) \right].  \nonumber
\end{align}

\subsection{Surface Area Scaling}

\begin{figure}
  \includegraphics[width=.75\linewidth, height=0.5\linewidth]{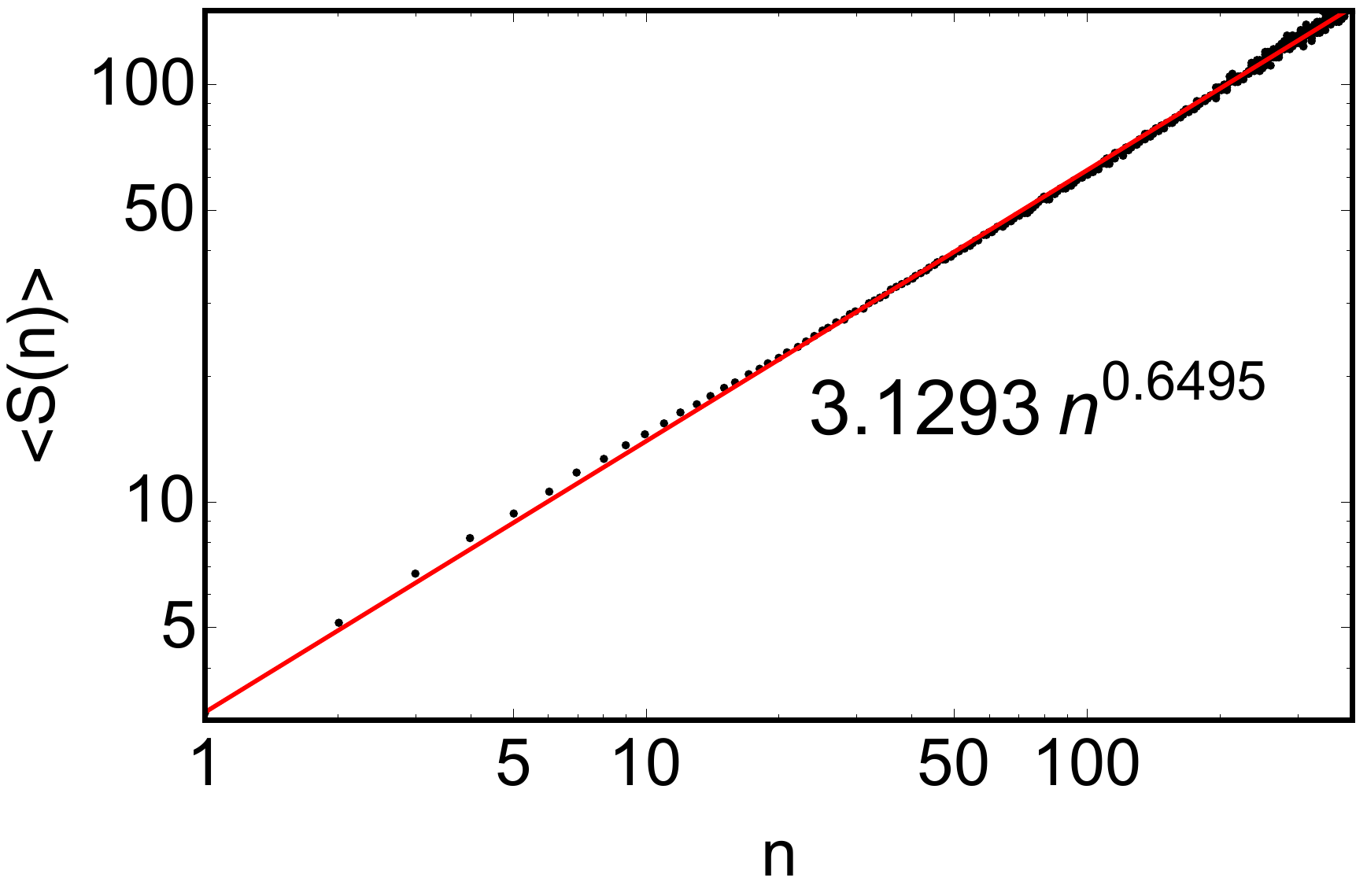}
\includegraphics[width=.75\linewidth, height=0.5\linewidth]{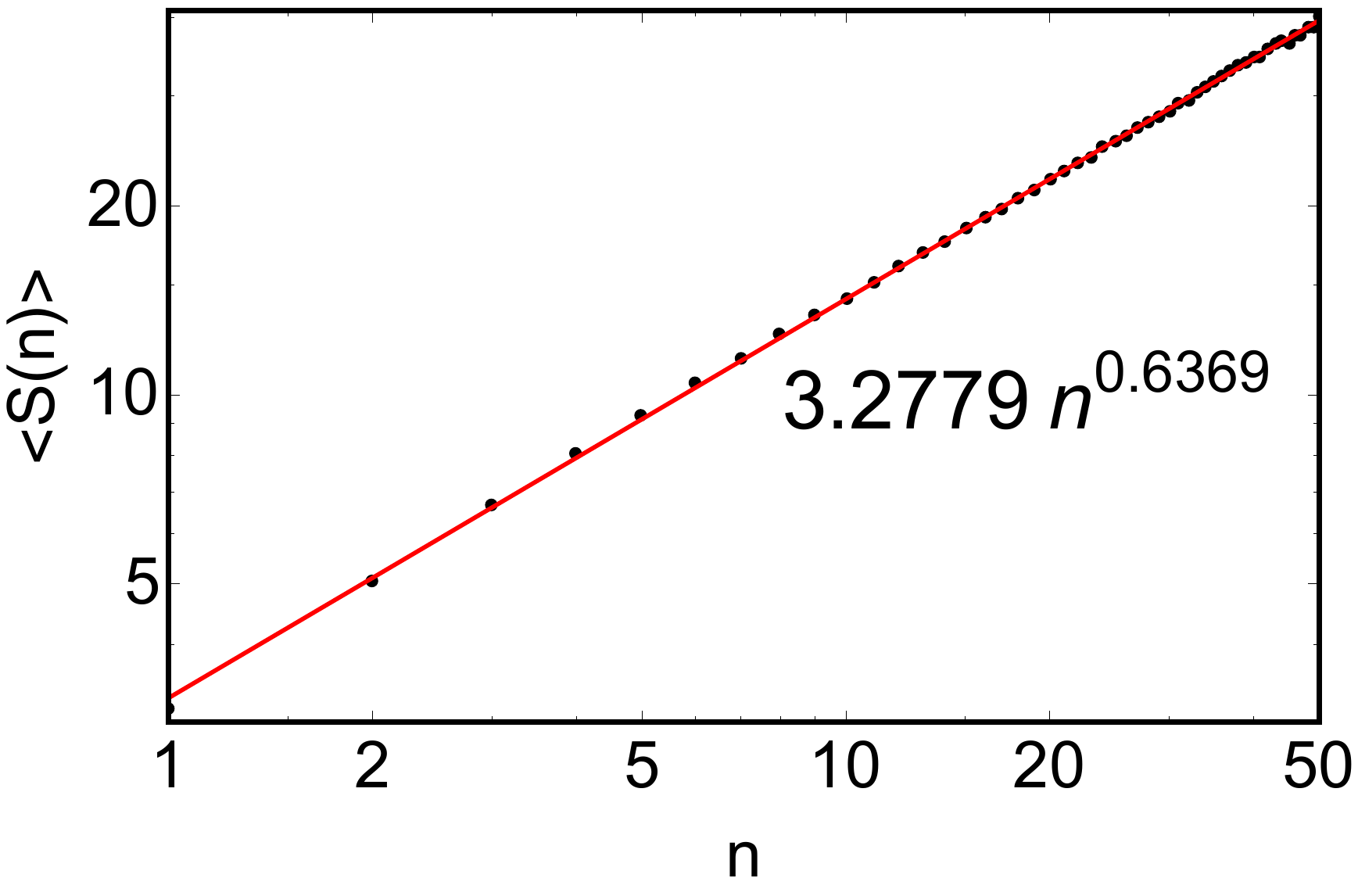}
  \caption{Two examples of the power law fits used to determine the surface area scaling. Red denotes the power law fit, and black disks the simulation data. Both plots were obtained from simulations on the binodal, with $(\vp, \phi) = (50, 0.502)$ [top] and $(\vp, \phi) = (100, 0.289)$ [bottom].}
  \label{fig:surfacearea_fits}
\end{figure}

We formalize the measure of cluster surface area using the notion of an $\alpha$-shape, which defines a concave hull of a set of points with respect to the spatial resolution $\alpha$ \cite{Edelsbrunner1983}.  By replacing each monomer with a constellation of points on its surface and taking $\alpha$ to be the monomer radius, the $\alpha$-shape corresponds to the physically relevant cluster surface.
We performed the calculation using the $\alpha$-shape algorithm from the Computational Geometry Algorithms Library (CGAL), on configurations from the simulations used to measure the cluster size distributions. The surface area for clusters of each size $n$ was averaged over the entire simulation, and the resulting distribution was fit to a power law (Fig.~\ref{fig:surfacearea_fits}). A random error of $\pm 0.0013$ was obtained by running 16 simulations at the same point in phase space ($\vp = 100, \phi = 0.29$) and taking the standard deviation of the measured scaling exponent. Somewhat larger variations of about $\pm 0.01$ were observed as $\vp$ and $\phi$ were varied independently, but without any obvious trend. Since such small variations are irrelevant for our purposes, we take for the scaling exponent the overall average $0.64 \pm 0.01$.

\begin{figure}
 \includegraphics[width=.49\linewidth, height=.49\linewidth]{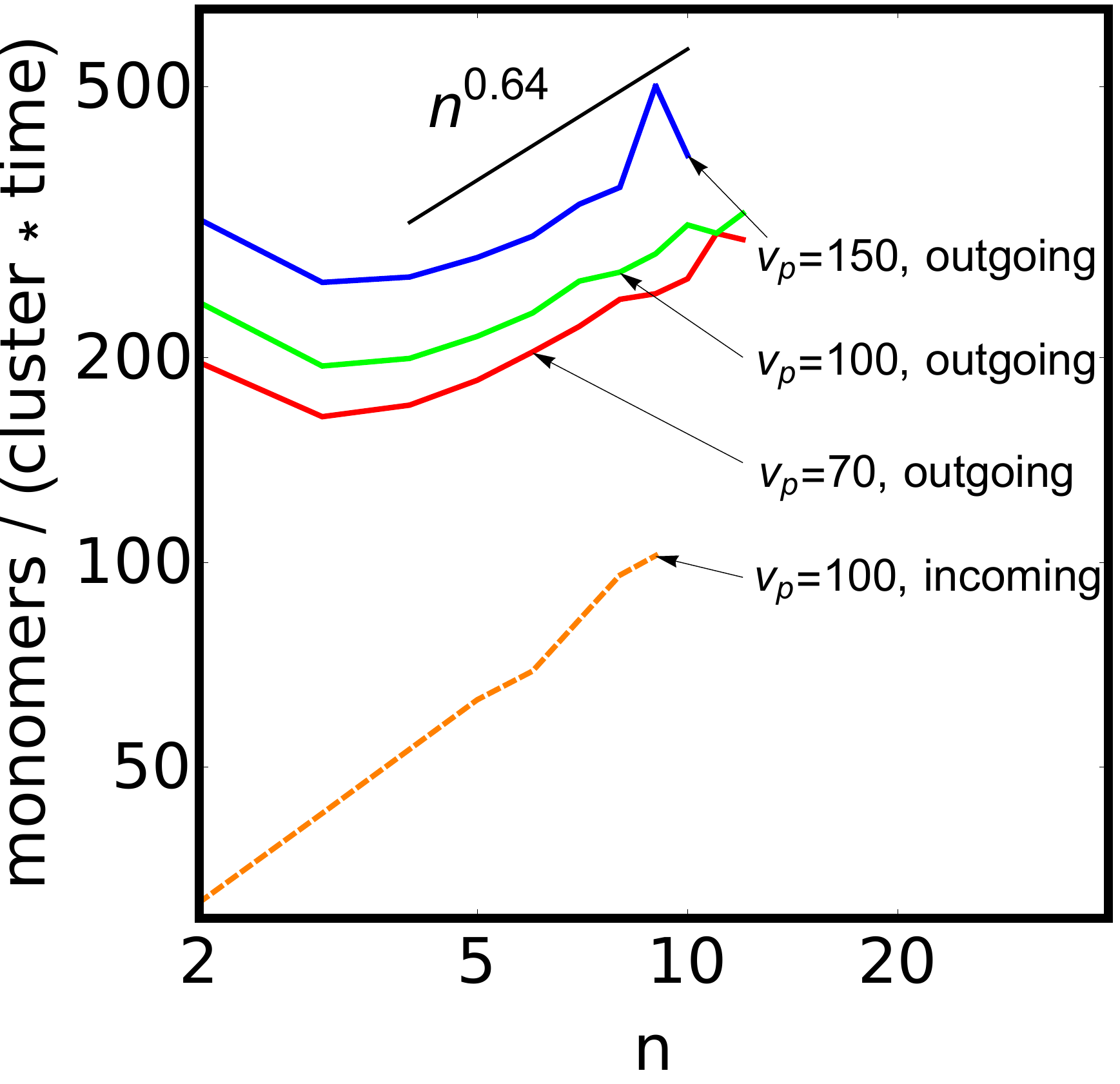}
\includegraphics[width=.49\linewidth, height=.49\linewidth]{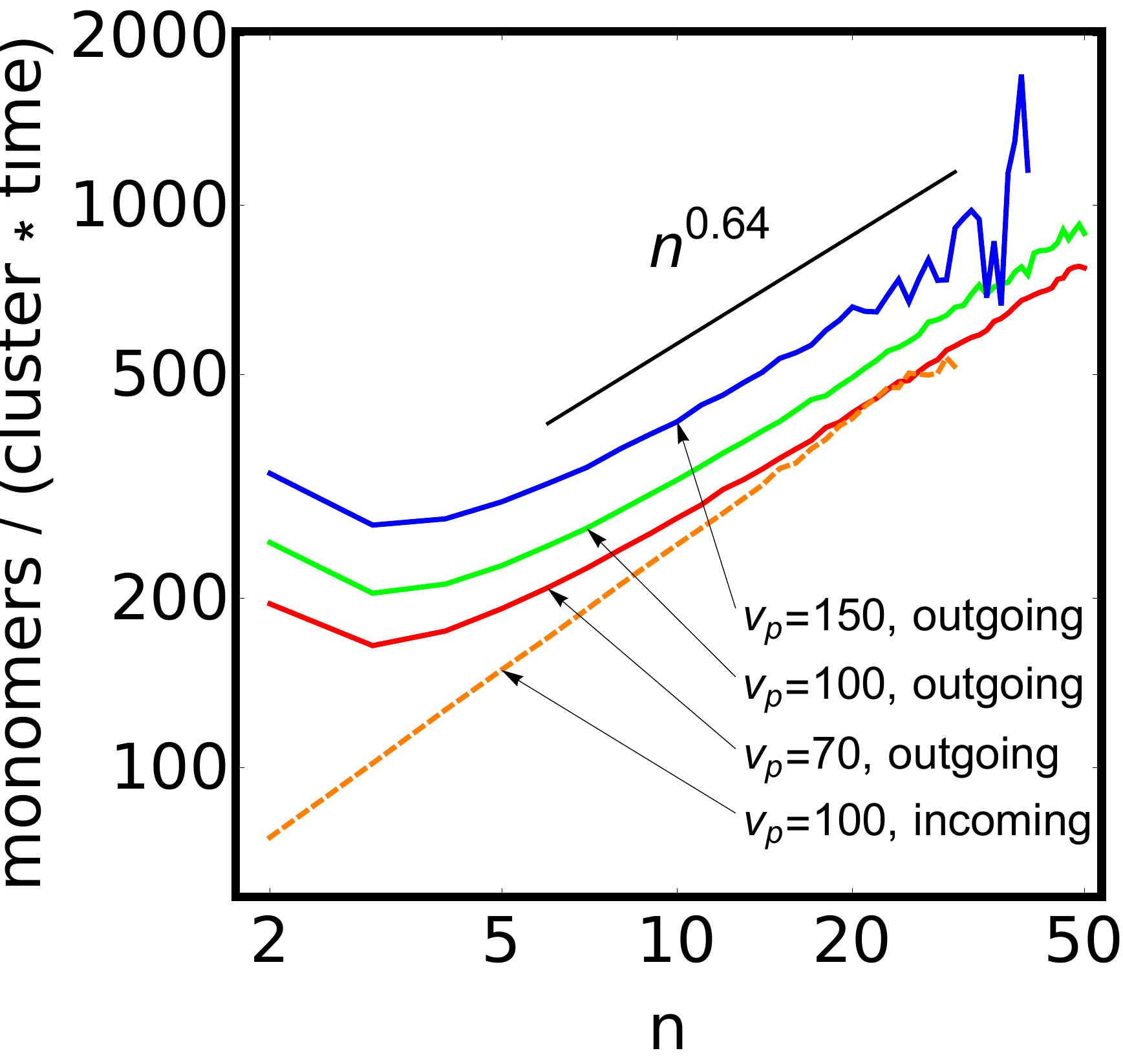}
\includegraphics[width=0.6\linewidth]{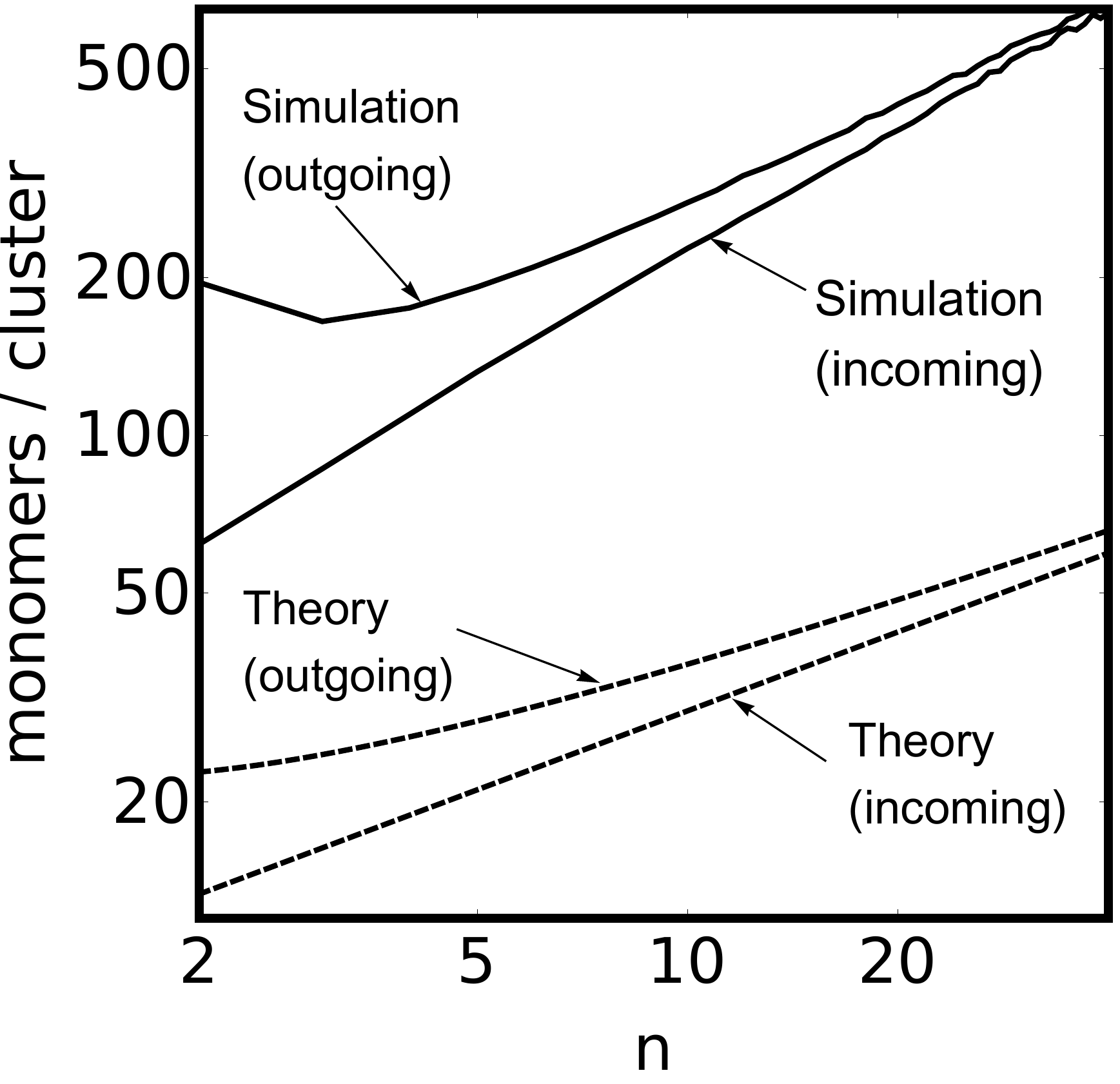}
  \caption{Top left: Comparison of incoming and outgoing rates for various values of $\vp$ below the binodal, with $z \rhog = 0.6$. The rates shown are integrated over the cluster surface and then averaged over the ensemble of clusters with size $n$; hence, the units are (monomers/(cluster * time)). Top right: Same as on the left, except on the binodal ($z \rhog = 1$).  Bottom: Comparison of rates on the binodal from simulations (solid black) and as computed by the kinetic theory (dashed black). The theoretical rates are integrated with respect to surface area scaling $S(n) = \pi n^{1/2}$.}
  \label{fig:rates}
\end{figure}

\begin{figure}
  \includegraphics[width=.75\linewidth]{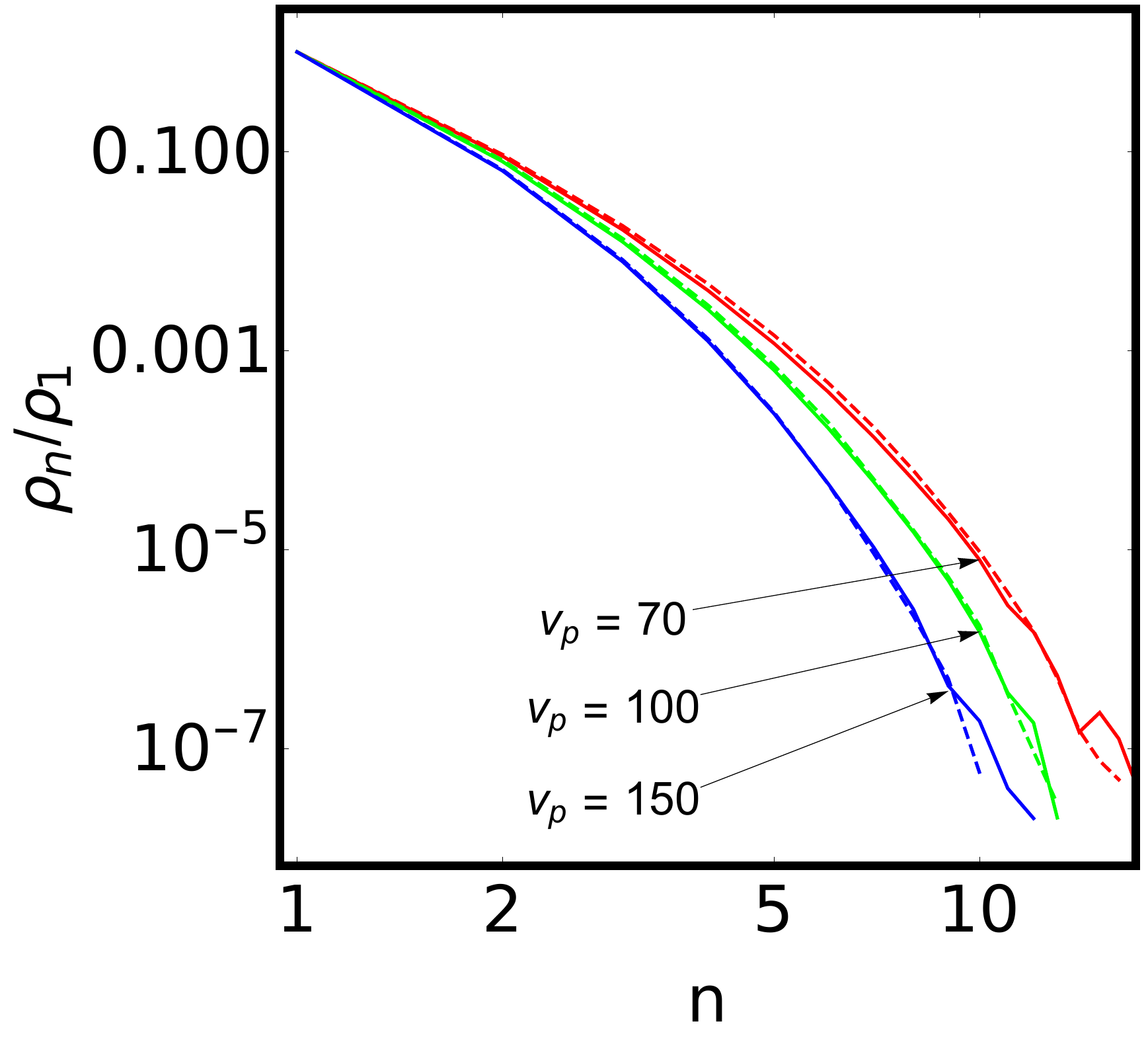}
\includegraphics[width=.75\linewidth]{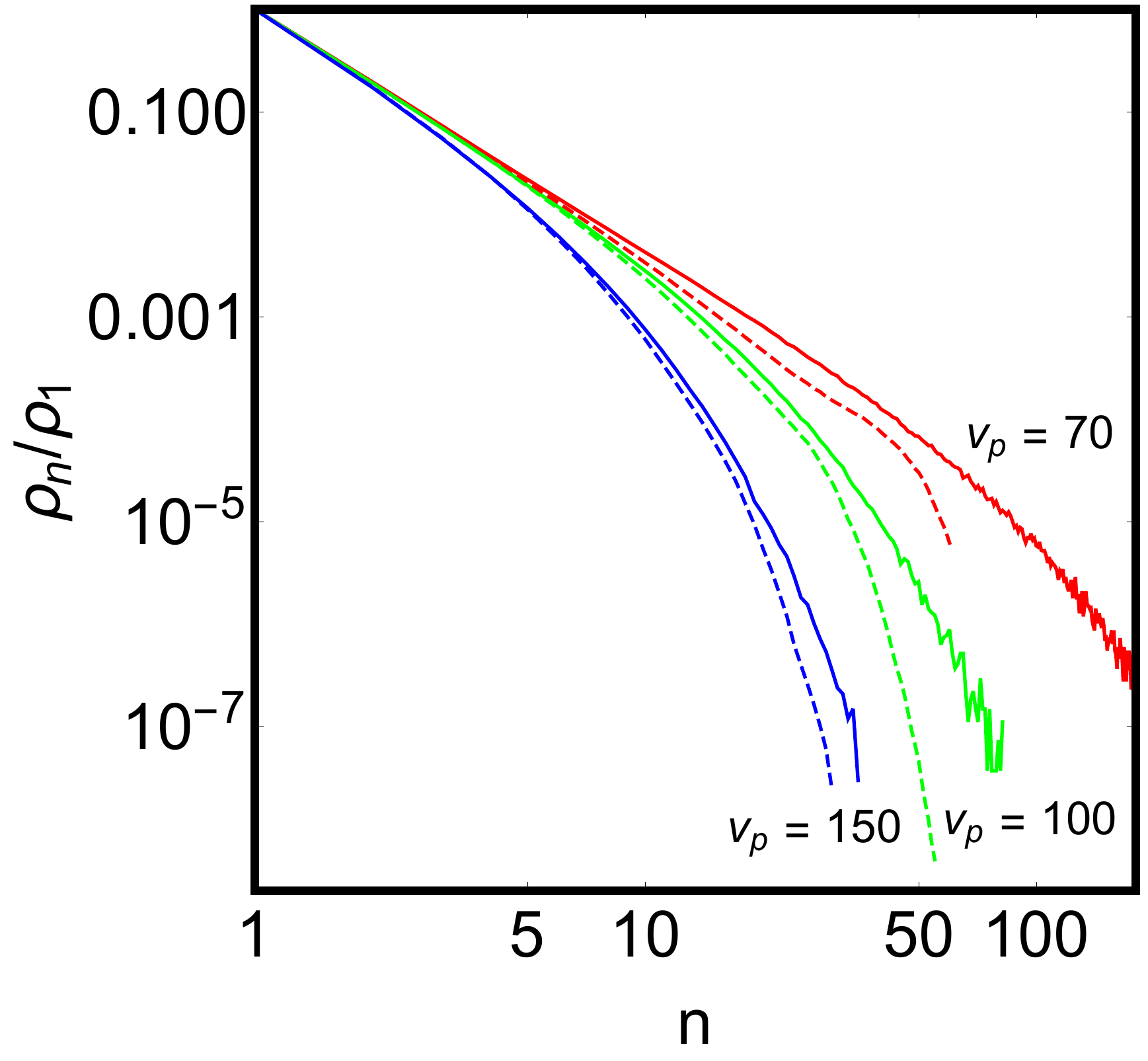}
  \caption{Top: Cluster size distributions (CSDs) below the binodal, with $z \rhog = 0.6$. The simulated CSDs are solid red, green, and blue. The CSDs determined by feeding directly measured rates into Eq. \ref{eq:model_cluster_size} are dashed red, green, and blue. Bottom: Same as on the top, except at the binodal ($z \rhog = 1$).}
  \label{fig:modified_CSDs}
\end{figure}

\subsection{Measurement of evaporation and adsorption rates}

To measure evaporation and adsorption rates, simulations were performed as described in Section I, but without translational diffusion in order to avoid counting events in which a monomer rapidly leaves and rejoins a cluster. This is justified since we are working in the regime where $\vp \sigma / D \gg 1$, such that any true evaporation or adsorption event is unlikely to be caused by translational diffusion. Each simulation was run until $10 \tau$, with the first $1 \tau$ of data discarded. To measure rates, a small time interval $\Delta t = 0.0001 \tau$ was considered, during which the probability of two evaporation or adsorption events occurring simultaneously on the same cluster is low. If a monomer was identified which $\Delta t$ units previously was located on the boundary of a cluster of size $n$, one evaporation event for that value of $n$ was counted. Conversely, if a cluster of size $n$ was found with $x$ monomers on its boundary that previously did not belong to a cluster, $x$ adsorption events were counted. The number of events was then summed to obtain average rates for each cluster size. Results are given in Fig. \ref{fig:rates} for $z \rhog = 0.6$ and $z \rhog = 1$ (the binodal), together with a direct comparison to the rates predicted by the kinetic theory on the binodal. Finally, the rates were inputted into the kinetic theory from the main text in order to reconstruct the CSDs. These results are presented in Fig. \ref{fig:modified_CSDs}, showing good agreement with simulation.

We note that the rates in Fig. \ref{fig:rates} are measured along iso-criticals $z \rhog = constant$, and thus $\vp$ and the overall density $\phi$ simultaneously change between the various curves shown. On the other hand, in the main text we have identified the primary contribution to the observed deviation of the outgoing rate from the simple theory as coming from $\vp$. This is \emph{prima facie} reasonable, since the outgoing rate is determined by the particles' interactions with the local geometry of a cluster surface, which simulations suggest does not change too radically as we vary $\phi$. To check this conclusion, we have measured outgoing rates holding $\phi$ and $\vp$ separately constant, and found that increasing $\phi$ while holding $\vp$ fixed only marginally increases the outgoing rates, whereas holding $\phi$ fixed while increasing $\vp$ causes the rates to increase by a significant amount. Thus, our identification of $\vp$ as the primary contributor is well justified.

{\bf Effects of cluster-cluster interactions.}
While the CSDs computed from the measured rates fit the simulation CSDs very well for small and moderate cluster sizes, they begin to deviate in the limit of large clusters. This deviation likely arises because, for sufficiently high system densities, the growth of large clusters  is no longer dominated by single monomer adsorption, as is assumed in our theory. More precisely, we must consider the general class of reaction pathways in which clusters of sizes $k$ and $n-k$ combine to form a cluster of size $n$, and $k$ may range from $1$ to $\lfloor n/2 \rfloor$. For large clusters in dense systems, it may no longer be reliable to ignore all reaction pathways in which $k \neq 1$. Thus, while not directly responsible for the appearance of power law scaling in the CSDs, cluster interactions appear to influence the location of the threshold size characterizing the crossover from power law scaling to exponential decay. Indeed, it is surprising that our model, which assumes non-interacting clusters, works as well as it does in the high density region.

%

\subsection{Derivation of the Coordinate Transformation}
By definition of the overall volume fraction $\phi$, we may write
\begin{align}
\phi = \frac{\pi \sigma ^{2}}{4}\sum_{n=1}^{n_{\max }}n\rho _{n}.
\label{b1}
\end{align}
In the single phase regime or at the binodal, the sum in (\ref{b1}) rigorously
should be extended to $\infty$. However, for purposes of numerical
calculation, a sufficiently accurate estimate is achieved for $n_{\max }=1000$.
In the metastable regime we assume that nucleation has not
yet occurred, such that terms with large enough $n$ do not contribute to
the sum. In practice, we enforce this constraint by setting $n_{\max
}=\min (n_\text{crit}, 1000)$. Inserting the expression for the cluster size distribution, we get
\begin{eqnarray}
\phi &=&\frac{\pi \sigma ^{2}}{4}\sum_{n=1}^{n_{\max }}n\rho_{1}P\left(
n\right)  \label{b2}
\label{b3}
\end{eqnarray}
To eliminate the unknown monomor density, $\rho _{1}$, we make use of the
definition of $\rhog$, which can be written as the number of monomers
divided by the `free volume' unoccupied by other clusters:%
\begin{align}
\rhog &=\frac{\rho _{1}V}{V-\sum_{n=2}^{n_{\max }}V\left( n\right)
\rho _{n}V} \\
&= \frac{\rho _{1}}{1-\rho _{1}\sum_{n=2}^{n_{max }}V(n)P(n)}
\end{align}
with $V$ the total system volume and $V(n)$ the volume of a cluster with size $n$.
Solving for $\rho_{1}$ and substituting into (\ref{b3}) gives
\begin{align}
\phi = \frac{\pi \sigma ^{2}}{4}\rho _{g}\sum_{n=1}^{n_{max }}nP(n)\left[ 1+\rho
_{g}\sum_{n=2}^{n_{max }}V(n)P(n)\right] ^{-1}
\label{b4}
\end{align}
Finally, using $V(n) = n/\rhoc$ and the definition of $P\left( n\right)$, this can be rearranged to read
\begin{align}
\phi = \frac{A \rhoc}{\left(4/\pi \sigma^2\right)A- 1 + \rhoc/\rhog}
\label{b5}
\end{align}
where $A=\frac{\pi \sigma^{2}}{4}\sum_{n=1}^{n_{max}} n P(n)$.


In the theory, we normally start by assuming values of $\vp$ and $z \rho_g$. These two quantities uniquely define a point in the usual $(\vp, \phi)$ space, with $z \rho_g$ also specifying the predicted phase behavior of the system (e.g. $z \rho_g = 1$ at the binodal). Given these quantities, we can solve for $\rho_g$ using the definition of $z$. Knowing $\rho_g$, we can then employ Eq. \ref{b5} to obtain $\phi$.

\bibliographystyle{apsrev4-1}
\bibliography{RednerDissertation}

\begin{thebibliography}{94}%
\makeatletter
\providecommand \@ifxundefined [1]{%
 \@ifx{#1\undefined}
}%
\providecommand \@ifnum [1]{%
 \ifnum #1\expandafter \@firstoftwo
 \else \expandafter \@secondoftwo
 \fi
}%
\providecommand \@ifx [1]{%
 \ifx #1\expandafter \@firstoftwo
 \else \expandafter \@secondoftwo
 \fi
}%
\providecommand \natexlab [1]{#1}%
\providecommand \enquote  [1]{``#1''}%
\providecommand \bibnamefont  [1]{#1}%
\providecommand \bibfnamefont [1]{#1}%
\providecommand \citenamefont [1]{#1}%
\providecommand \href@noop [0]{\@secondoftwo}%
\providecommand \href [0]{\begingroup \@sanitize@url \@href}%
\providecommand \@href[1]{\@@startlink{#1}\@@href}%
\providecommand \@@href[1]{\endgroup#1\@@endlink}%
\providecommand \@sanitize@url [0]{\catcode `\\12\catcode `\$12\catcode
  `\&12\catcode `\#12\catcode `\^12\catcode `\_12\catcode `\%12\relax}%
\providecommand \@@startlink[1]{}%
\providecommand \@@endlink[0]{}%
\providecommand \url  [0]{\begingroup\@sanitize@url \@url }%
\providecommand \@url [1]{\endgroup\@href {#1}{\urlprefix }}%
\providecommand \urlprefix  [0]{URL }%
\providecommand \Eprint [0]{\href }%
\providecommand \doibase [0]{http://dx.doi.org/}%
\providecommand \selectlanguage [0]{\@gobble}%
\providecommand \bibinfo  [0]{\@secondoftwo}%
\providecommand \bibfield  [0]{\@secondoftwo}%
\providecommand \translation [1]{[#1]}%
\providecommand \BibitemOpen [0]{}%
\providecommand \bibitemStop [0]{}%
\providecommand \bibitemNoStop [0]{.\EOS\space}%
\providecommand \EOS [0]{\spacefactor3000\relax}%
\providecommand \BibitemShut  [1]{\csname bibitem#1\endcsname}%
\let\auto@bib@innerbib\@empty
\bibitem [{\citenamefont {Brugu\'{e}s}\ \emph {et~al.}(2012)\citenamefont
  {Brugu\'{e}s}, \citenamefont {Nuzzo}, \citenamefont {Mazur},\ and\
  \citenamefont {Needleman}}]{Brugues2012}%
  \BibitemOpen
  \bibfield  {author} {\bibinfo {author} {\bibfnamefont {J.}~\bibnamefont
  {Brugu\'{e}s}}, \bibinfo {author} {\bibfnamefont {V.}~\bibnamefont {Nuzzo}},
  \bibinfo {author} {\bibfnamefont {E.}~\bibnamefont {Mazur}}, \ and\ \bibinfo
  {author} {\bibfnamefont {D.~J.}\ \bibnamefont {Needleman}},\ }\href {\doibase
  10.1016/j.cell.2012.03.027} {\bibfield  {journal} {\bibinfo  {journal}
  {Cell}\ }\textbf {\bibinfo {volume} {149}},\ \bibinfo {pages} {554} (\bibinfo
  {year} {2012})}\BibitemShut {NoStop}%
\bibitem [{\citenamefont {Brugues}\ and\ \citenamefont
  {Needleman}(2014)}]{Brugues2014}%
  \BibitemOpen
  \bibfield  {author} {\bibinfo {author} {\bibfnamefont {J.}~\bibnamefont
  {Brugues}}\ and\ \bibinfo {author} {\bibfnamefont {D.}~\bibnamefont
  {Needleman}},\ }\href {\doibase 10.1073/pnas.1409404111} {\bibfield
  {journal} {\bibinfo  {journal} {Proc. Natl. Acad. Sci. U. S. A.}\ }\textbf
  {\bibinfo {volume} {111}},\ \bibinfo {pages} {18496} (\bibinfo {year}
  {2014})}\BibitemShut {NoStop}%
\bibitem [{\citenamefont {Goldstein}\ and\ \citenamefont {van~de
  Meent}(2015)}]{Goldstein2015}%
  \BibitemOpen
  \bibfield  {author} {\bibinfo {author} {\bibfnamefont {R.~E.}\ \bibnamefont
  {Goldstein}}\ and\ \bibinfo {author} {\bibfnamefont {J.-W.}\ \bibnamefont
  {van~de Meent}},\ }\href {\doibase 10.1098/rsfs.2015.0030} {\bibfield
  {journal} {\bibinfo  {journal} {Interface Focus}\ }\textbf {\bibinfo {volume}
  {5}},\ \bibinfo {pages} {20150030} (\bibinfo {year} {2015})}\BibitemShut
  {NoStop}%
\bibitem [{\citenamefont {Cisneros}\ \emph {et~al.}(2011)\citenamefont
  {Cisneros}, \citenamefont {Kessler}, \citenamefont {Ganguly},\ and\
  \citenamefont {Goldstein}}]{Cisneros2011a}%
  \BibitemOpen
  \bibfield  {author} {\bibinfo {author} {\bibfnamefont {L.}~\bibnamefont
  {Cisneros}}, \bibinfo {author} {\bibfnamefont {J.}~\bibnamefont {Kessler}},
  \bibinfo {author} {\bibfnamefont {S.}~\bibnamefont {Ganguly}}, \ and\
  \bibinfo {author} {\bibfnamefont {R.}~\bibnamefont {Goldstein}},\ }\href
  {\doibase 10.1103/PhysRevE.83.061907} {\bibfield  {journal} {\bibinfo
  {journal} {Phys. Rev. E}\ }\textbf {\bibinfo {volume} {83}},\ \bibinfo
  {pages} {061907} (\bibinfo {year} {2011})}\BibitemShut {NoStop}%
\bibitem [{\citenamefont {Dombrowski}\ \emph {et~al.}(2004)\citenamefont
  {Dombrowski}, \citenamefont {Cisneros}, \citenamefont {Chatkaew},
  \citenamefont {Goldstein},\ and\ \citenamefont {Kessler}}]{Dombrowski2004}%
  \BibitemOpen
  \bibfield  {author} {\bibinfo {author} {\bibfnamefont {C.}~\bibnamefont
  {Dombrowski}}, \bibinfo {author} {\bibfnamefont {L.}~\bibnamefont
  {Cisneros}}, \bibinfo {author} {\bibfnamefont {S.}~\bibnamefont {Chatkaew}},
  \bibinfo {author} {\bibfnamefont {R.~E.}\ \bibnamefont {Goldstein}}, \ and\
  \bibinfo {author} {\bibfnamefont {J.~O.}\ \bibnamefont {Kessler}},\
  }\href@noop {} {\bibfield  {journal} {\bibinfo  {journal} {Phys. Rev. Lett.}\
  }\textbf {\bibinfo {volume} {93}},\ \bibinfo {pages} {098103 (4 pages)}
  (\bibinfo {year} {2004})}\BibitemShut {NoStop}%
\bibitem [{\citenamefont {Kaiser}\ \emph {et~al.}(2014)\citenamefont {Kaiser},
  \citenamefont {Peshkov}, \citenamefont {Sokolov}, \citenamefont {ten Hagen},
  \citenamefont {L�wen},\ and\ \citenamefont {Aranson}}]{Kaiser2014}%
  \BibitemOpen
  \bibfield  {author} {\bibinfo {author} {\bibfnamefont {A.}~\bibnamefont
  {Kaiser}}, \bibinfo {author} {\bibfnamefont {A.}~\bibnamefont {Peshkov}},
  \bibinfo {author} {\bibfnamefont {A.}~\bibnamefont {Sokolov}}, \bibinfo
  {author} {\bibfnamefont {B.}~\bibnamefont {ten Hagen}}, \bibinfo {author}
  {\bibfnamefont {H.}~\bibnamefont {L�wen}}, \ and\ \bibinfo {author}
  {\bibfnamefont {I.~S.}\ \bibnamefont {Aranson}},\ }\href
  {http://link.aps.org/doi/10.1103/PhysRevLett.112.158101} {\bibfield
  {journal} {\bibinfo  {journal} {Phys. Rev. Lett.}\ }\textbf {\bibinfo
  {volume} {112}},\ \bibinfo {pages} {158101} (\bibinfo {year}
  {2014})}\BibitemShut {NoStop}%
\bibitem [{\citenamefont {Dunkel}\ \emph {et~al.}(2013)\citenamefont {Dunkel},
  \citenamefont {Heidenreich}, \citenamefont {Drescher}, \citenamefont
  {Wensink}, \citenamefont {B\"{a}r},\ and\ \citenamefont
  {Goldstein}}]{Dunkel2013}%
  \BibitemOpen
  \bibfield  {author} {\bibinfo {author} {\bibfnamefont {J.}~\bibnamefont
  {Dunkel}}, \bibinfo {author} {\bibfnamefont {S.}~\bibnamefont {Heidenreich}},
  \bibinfo {author} {\bibfnamefont {K.}~\bibnamefont {Drescher}}, \bibinfo
  {author} {\bibfnamefont {H.~H.}\ \bibnamefont {Wensink}}, \bibinfo {author}
  {\bibfnamefont {M.}~\bibnamefont {B\"{a}r}}, \ and\ \bibinfo {author}
  {\bibfnamefont {R.~E.}\ \bibnamefont {Goldstein}},\ }\href {\doibase
  10.1103/PhysRevLett.110.228102} {\bibfield  {journal} {\bibinfo  {journal}
  {Phys. Rev. Lett.}\ }\textbf {\bibinfo {volume} {110}},\ \bibinfo {pages}
  {228102} (\bibinfo {year} {2013})}\BibitemShut {NoStop}%
\bibitem [{\citenamefont {Attanasi}\ \emph
  {et~al.}(2014{\natexlab{a}})\citenamefont {Attanasi}, \citenamefont
  {Cavagna}, \citenamefont {Castello}, \citenamefont {Giardina}, \citenamefont
  {Grigera}, \citenamefont {Jeli\'{c}}, \citenamefont {Melillo}, \citenamefont
  {Parisi}, \citenamefont {Pohl}, \citenamefont {Shen},\ and\ \citenamefont
  {Viale}}]{Attanasi2014b}%
  \BibitemOpen
  \bibfield  {author} {\bibinfo {author} {\bibfnamefont {A.}~\bibnamefont
  {Attanasi}}, \bibinfo {author} {\bibfnamefont {A.}~\bibnamefont {Cavagna}},
  \bibinfo {author} {\bibfnamefont {L.~D.}\ \bibnamefont {Castello}}, \bibinfo
  {author} {\bibfnamefont {I.}~\bibnamefont {Giardina}}, \bibinfo {author}
  {\bibfnamefont {T.~S.}\ \bibnamefont {Grigera}}, \bibinfo {author}
  {\bibfnamefont {A.}~\bibnamefont {Jeli\'{c}}}, \bibinfo {author}
  {\bibfnamefont {S.}~\bibnamefont {Melillo}}, \bibinfo {author} {\bibfnamefont
  {L.}~\bibnamefont {Parisi}}, \bibinfo {author} {\bibfnamefont
  {O.}~\bibnamefont {Pohl}}, \bibinfo {author} {\bibfnamefont {E.}~\bibnamefont
  {Shen}}, \ and\ \bibinfo {author} {\bibfnamefont {M.}~\bibnamefont {Viale}},\
  }\href {\doibase 10.1038/NPHYS3035} {\bibfield  {journal} {\bibinfo
  {journal} {Nat. Phys.}\ }\textbf {\bibinfo {volume} {10}},\ \bibinfo {pages}
  {691} (\bibinfo {year} {2014}{\natexlab{a}})}\BibitemShut {NoStop}%
\bibitem [{\citenamefont {Attanasi}\ \emph
  {et~al.}(2014{\natexlab{b}})\citenamefont {Attanasi}, \citenamefont
  {Cavagna}, \citenamefont {{Del Castello}}, \citenamefont {Giardina},
  \citenamefont {Melillo}, \citenamefont {Parisi}, \citenamefont {Pohl},
  \citenamefont {Rossaro}, \citenamefont {Shen}, \citenamefont {Silvestri},\
  and\ \citenamefont {Viale}}]{Attanasi2014c}%
  \BibitemOpen
  \bibfield  {author} {\bibinfo {author} {\bibfnamefont {A.}~\bibnamefont
  {Attanasi}}, \bibinfo {author} {\bibfnamefont {A.}~\bibnamefont {Cavagna}},
  \bibinfo {author} {\bibfnamefont {L.}~\bibnamefont {{Del Castello}}},
  \bibinfo {author} {\bibfnamefont {I.}~\bibnamefont {Giardina}}, \bibinfo
  {author} {\bibfnamefont {S.}~\bibnamefont {Melillo}}, \bibinfo {author}
  {\bibfnamefont {L.}~\bibnamefont {Parisi}}, \bibinfo {author} {\bibfnamefont
  {O.}~\bibnamefont {Pohl}}, \bibinfo {author} {\bibfnamefont {B.}~\bibnamefont
  {Rossaro}}, \bibinfo {author} {\bibfnamefont {E.}~\bibnamefont {Shen}},
  \bibinfo {author} {\bibfnamefont {E.}~\bibnamefont {Silvestri}}, \ and\
  \bibinfo {author} {\bibfnamefont {M.}~\bibnamefont {Viale}},\ }\href
  {\doibase 10.1371/journal.pcbi.1003697} {\bibfield  {journal} {\bibinfo
  {journal} {PLoS Comput. Biol.}\ }\textbf {\bibinfo {volume} {10}},\ \bibinfo
  {pages} {e1003697} (\bibinfo {year} {2014}{\natexlab{b}})}\BibitemShut
  {NoStop}%
\bibitem [{\citenamefont {Shaw}(1978)}]{Shaw1978}%
  \BibitemOpen
  \bibfield  {author} {\bibinfo {author} {\bibfnamefont {E.}~\bibnamefont
  {Shaw}},\ }\href@noop {} {\bibfield  {journal} {\bibinfo  {journal} {Amer.
  Sci.}\ }\textbf {\bibinfo {volume} {66}},\ \bibinfo {pages} {166} (\bibinfo
  {year} {1978})}\BibitemShut {NoStop}%
\bibitem [{\citenamefont {H\"{o}lldobler}\ and\ \citenamefont
  {Wilson}(1994)}]{Holldobler1994}%
  \BibitemOpen
  \bibfield  {author} {\bibinfo {author} {\bibfnamefont {B.}~\bibnamefont
  {H\"{o}lldobler}}\ and\ \bibinfo {author} {\bibfnamefont {E.~O.}\
  \bibnamefont {Wilson}},\ }\href@noop {} {\emph {\bibinfo {title} {{Journey to
  the Ants: A Story of Scientific Exploration}}}}\ (\bibinfo  {publisher}
  {Belknap Press of Harvard University Press},\ \bibinfo {year}
  {1994})\BibitemShut {NoStop}%
\bibitem [{\citenamefont {Palacci}\ \emph {et~al.}(2010)\citenamefont
  {Palacci}, \citenamefont {Cottin-Bizonne}, \citenamefont {Ybert},\ and\
  \citenamefont {Bocquet}}]{Palacci2010a}%
  \BibitemOpen
  \bibfield  {author} {\bibinfo {author} {\bibfnamefont {J.}~\bibnamefont
  {Palacci}}, \bibinfo {author} {\bibfnamefont {C.}~\bibnamefont
  {Cottin-Bizonne}}, \bibinfo {author} {\bibfnamefont {C.}~\bibnamefont
  {Ybert}}, \ and\ \bibinfo {author} {\bibfnamefont {L.}~\bibnamefont
  {Bocquet}},\ }\href {\doibase 10.1103/PhysRevLett.105.088304} {\bibfield
  {journal} {\bibinfo  {journal} {Phys. Rev. Lett.}\ }\textbf {\bibinfo
  {volume} {105}},\ \bibinfo {pages} {088304} (\bibinfo {year}
  {2010})}\BibitemShut {NoStop}%
\bibitem [{\citenamefont {Paxton}\ \emph {et~al.}(2004)\citenamefont {Paxton},
  \citenamefont {Kistler}, \citenamefont {Olmeda}, \citenamefont {Sen},
  \citenamefont {{St Angelo}}, \citenamefont {Cao}, \citenamefont {Mallouk},
  \citenamefont {Lammert},\ and\ \citenamefont {Crespi}}]{Paxton2004}%
  \BibitemOpen
  \bibfield  {author} {\bibinfo {author} {\bibfnamefont {W.~F.}\ \bibnamefont
  {Paxton}}, \bibinfo {author} {\bibfnamefont {K.~C.}\ \bibnamefont {Kistler}},
  \bibinfo {author} {\bibfnamefont {C.~C.}\ \bibnamefont {Olmeda}}, \bibinfo
  {author} {\bibfnamefont {A.}~\bibnamefont {Sen}}, \bibinfo {author}
  {\bibfnamefont {S.~K.}\ \bibnamefont {{St Angelo}}}, \bibinfo {author}
  {\bibfnamefont {Y.}~\bibnamefont {Cao}}, \bibinfo {author} {\bibfnamefont
  {T.~E.}\ \bibnamefont {Mallouk}}, \bibinfo {author} {\bibfnamefont {P.~E.}\
  \bibnamefont {Lammert}}, \ and\ \bibinfo {author} {\bibfnamefont {V.~H.}\
  \bibnamefont {Crespi}},\ }\href {\doibase 10.1021/ja047697z} {\bibfield
  {journal} {\bibinfo  {journal} {J. Am. Chem. Soc.}\ }\textbf {\bibinfo
  {volume} {126}},\ \bibinfo {pages} {13424} (\bibinfo {year}
  {2004})}\BibitemShut {NoStop}%
\bibitem [{\citenamefont {Hong}\ \emph {et~al.}(2007)\citenamefont {Hong},
  \citenamefont {Blackman}, \citenamefont {Kopp}, \citenamefont {Sen},\ and\
  \citenamefont {Velegol}}]{Hong2007}%
  \BibitemOpen
  \bibfield  {author} {\bibinfo {author} {\bibfnamefont {Y.}~\bibnamefont
  {Hong}}, \bibinfo {author} {\bibfnamefont {N.}~\bibnamefont {Blackman}},
  \bibinfo {author} {\bibfnamefont {N.}~\bibnamefont {Kopp}}, \bibinfo {author}
  {\bibfnamefont {A.}~\bibnamefont {Sen}}, \ and\ \bibinfo {author}
  {\bibfnamefont {D.}~\bibnamefont {Velegol}},\ }\href {\doibase
  10.1103/PhysRevLett.99.178103} {\bibfield  {journal} {\bibinfo  {journal}
  {Phys. Rev. Lett.}\ }\textbf {\bibinfo {volume} {99}},\ \bibinfo {pages}
  {178103} (\bibinfo {year} {2007})}\BibitemShut {NoStop}%
\bibitem [{\citenamefont {Jiang}\ \emph {et~al.}(2010)\citenamefont {Jiang},
  \citenamefont {Yoshinaga},\ and\ \citenamefont {Sano}}]{Jiang2010}%
  \BibitemOpen
  \bibfield  {author} {\bibinfo {author} {\bibfnamefont {H.-R.}\ \bibnamefont
  {Jiang}}, \bibinfo {author} {\bibfnamefont {N.}~\bibnamefont {Yoshinaga}}, \
  and\ \bibinfo {author} {\bibfnamefont {M.}~\bibnamefont {Sano}},\ }\href
  {\doibase 10.1103/PhysRevLett.105.268302} {\bibfield  {journal} {\bibinfo
  {journal} {Phys. Rev. Lett.}\ }\textbf {\bibinfo {volume} {105}},\ \bibinfo
  {pages} {268302} (\bibinfo {year} {2010})}\BibitemShut {NoStop}%
\bibitem [{\citenamefont {Volpe}\ \emph {et~al.}(2011)\citenamefont {Volpe},
  \citenamefont {Buttinoni}, \citenamefont {Vogt}, \citenamefont
  {K\"{u}mmerer},\ and\ \citenamefont {Bechinger}}]{Volpe2011}%
  \BibitemOpen
  \bibfield  {author} {\bibinfo {author} {\bibfnamefont {G.}~\bibnamefont
  {Volpe}}, \bibinfo {author} {\bibfnamefont {I.}~\bibnamefont {Buttinoni}},
  \bibinfo {author} {\bibfnamefont {D.}~\bibnamefont {Vogt}}, \bibinfo {author}
  {\bibfnamefont {H.-J.}\ \bibnamefont {K\"{u}mmerer}}, \ and\ \bibinfo
  {author} {\bibfnamefont {C.}~\bibnamefont {Bechinger}},\ }\href {\doibase
  10.1039/c1sm05960b} {\bibfield  {journal} {\bibinfo  {journal} {Soft Matter}\
  }\textbf {\bibinfo {volume} {7}},\ \bibinfo {pages} {8810} (\bibinfo {year}
  {2011})}\BibitemShut {NoStop}%
\bibitem [{\citenamefont {Thutupalli}\ \emph {et~al.}(2011)\citenamefont
  {Thutupalli}, \citenamefont {Seemann},\ and\ \citenamefont
  {Herminghaus}}]{Thutupalli2011}%
  \BibitemOpen
  \bibfield  {author} {\bibinfo {author} {\bibfnamefont {S.}~\bibnamefont
  {Thutupalli}}, \bibinfo {author} {\bibfnamefont {R.}~\bibnamefont {Seemann}},
  \ and\ \bibinfo {author} {\bibfnamefont {S.}~\bibnamefont {Herminghaus}},\
  }\href {\doibase 10.1088/1367-2630/13/7/073021} {\bibfield  {journal}
  {\bibinfo  {journal} {New Journal of Physics}\ }\textbf {\bibinfo {volume}
  {13}},\ \bibinfo {pages} {073021} (\bibinfo {year} {2011})}\BibitemShut
  {NoStop}%
\bibitem [{\citenamefont {Theurkauff}\ \emph {et~al.}(2012)\citenamefont
  {Theurkauff}, \citenamefont {Cottin-Bizonne}, \citenamefont {Palacci},
  \citenamefont {Ybert},\ and\ \citenamefont {Bocquet}}]{Theurkauff2012}%
  \BibitemOpen
  \bibfield  {author} {\bibinfo {author} {\bibfnamefont {I.}~\bibnamefont
  {Theurkauff}}, \bibinfo {author} {\bibfnamefont {C.}~\bibnamefont
  {Cottin-Bizonne}}, \bibinfo {author} {\bibfnamefont {J.}~\bibnamefont
  {Palacci}}, \bibinfo {author} {\bibfnamefont {C.}~\bibnamefont {Ybert}}, \
  and\ \bibinfo {author} {\bibfnamefont {L.}~\bibnamefont {Bocquet}},\ }\href
  {\doibase 10.1103/PhysRevLett.108.268303} {\bibfield  {journal} {\bibinfo
  {journal} {Phys. Rev. Lett.}\ }\textbf {\bibinfo {volume} {108}},\ \bibinfo
  {pages} {268303} (\bibinfo {year} {2012})}\BibitemShut {NoStop}%
\bibitem [{\citenamefont {Palacci}\ \emph {et~al.}(2013)\citenamefont
  {Palacci}, \citenamefont {Sacanna}, \citenamefont {Steinberg}, \citenamefont
  {Pine},\ and\ \citenamefont {Chaikin}}]{Palacci2013a}%
  \BibitemOpen
  \bibfield  {author} {\bibinfo {author} {\bibfnamefont {J.}~\bibnamefont
  {Palacci}}, \bibinfo {author} {\bibfnamefont {S.}~\bibnamefont {Sacanna}},
  \bibinfo {author} {\bibfnamefont {A.~P.}\ \bibnamefont {Steinberg}}, \bibinfo
  {author} {\bibfnamefont {D.~J.}\ \bibnamefont {Pine}}, \ and\ \bibinfo
  {author} {\bibfnamefont {P.~M.}\ \bibnamefont {Chaikin}},\ }\href {\doibase
  10.1126/science.1230020} {\bibfield  {journal} {\bibinfo  {journal}
  {Science}\ }\textbf {\bibinfo {volume} {339}},\ \bibinfo {pages} {936}
  (\bibinfo {year} {2013})}\BibitemShut {NoStop}%
\bibitem [{\citenamefont {Palacci}\ \emph {et~al.}(2014)\citenamefont
  {Palacci}, \citenamefont {Sacanna}, \citenamefont {Kim}, \citenamefont {Yi},
  \citenamefont {Pine},\ and\ \citenamefont {Chaikin}}]{Palacci2014}%
  \BibitemOpen
  \bibfield  {author} {\bibinfo {author} {\bibfnamefont {J.}~\bibnamefont
  {Palacci}}, \bibinfo {author} {\bibfnamefont {S.}~\bibnamefont {Sacanna}},
  \bibinfo {author} {\bibfnamefont {S.-H.}\ \bibnamefont {Kim}}, \bibinfo
  {author} {\bibfnamefont {G.-R.}\ \bibnamefont {Yi}}, \bibinfo {author}
  {\bibfnamefont {D.~J.}\ \bibnamefont {Pine}}, \ and\ \bibinfo {author}
  {\bibfnamefont {P.~M.}\ \bibnamefont {Chaikin}},\ }\href {\doibase
  10.1098/rsta.2013.0372} {\bibfield  {journal} {\bibinfo  {journal} {Phil.
  Trans. R. Soc. A}\ }\textbf {\bibinfo {volume} {372}},\ \bibinfo {pages}
  {20130372} (\bibinfo {year} {2014})}\BibitemShut {NoStop}%
\bibitem [{\citenamefont {Bricard}\ \emph {et~al.}(2013)\citenamefont
  {Bricard}, \citenamefont {Caussin}, \citenamefont {Desreumaux}, \citenamefont
  {Dauchot},\ and\ \citenamefont {Bartolo}}]{Bricard2013}%
  \BibitemOpen
  \bibfield  {author} {\bibinfo {author} {\bibfnamefont {A.}~\bibnamefont
  {Bricard}}, \bibinfo {author} {\bibfnamefont {J.-B.}\ \bibnamefont
  {Caussin}}, \bibinfo {author} {\bibfnamefont {N.}~\bibnamefont {Desreumaux}},
  \bibinfo {author} {\bibfnamefont {O.}~\bibnamefont {Dauchot}}, \ and\
  \bibinfo {author} {\bibfnamefont {D.}~\bibnamefont {Bartolo}},\ }\href
  {\doibase 10.1038/nature12673} {\bibfield  {journal} {\bibinfo  {journal}
  {Nature}\ }\textbf {\bibinfo {volume} {503}},\ \bibinfo {pages} {95}
  (\bibinfo {year} {2013})}\BibitemShut {NoStop}%
\bibitem [{\citenamefont {Narayan}\ \emph {et~al.}(2007)\citenamefont
  {Narayan}, \citenamefont {Ramaswamy},\ and\ \citenamefont
  {Menon}}]{Narayan2007}%
  \BibitemOpen
  \bibfield  {author} {\bibinfo {author} {\bibfnamefont {V.}~\bibnamefont
  {Narayan}}, \bibinfo {author} {\bibfnamefont {S.}~\bibnamefont {Ramaswamy}},
  \ and\ \bibinfo {author} {\bibfnamefont {N.}~\bibnamefont {Menon}},\ }\href
  {\doibase 10.1126/science.1140414} {\bibfield  {journal} {\bibinfo  {journal}
  {Science}\ }\textbf {\bibinfo {volume} {317}},\ \bibinfo {pages} {105}
  (\bibinfo {year} {2007})}\BibitemShut {NoStop}%
\bibitem [{\citenamefont {Kudrolli}\ \emph {et~al.}(2008)\citenamefont
  {Kudrolli}, \citenamefont {Lumay}, \citenamefont {Volfson},\ and\
  \citenamefont {Tsimring}}]{Kudrolli2008}%
  \BibitemOpen
  \bibfield  {author} {\bibinfo {author} {\bibfnamefont {A.}~\bibnamefont
  {Kudrolli}}, \bibinfo {author} {\bibfnamefont {G.}~\bibnamefont {Lumay}},
  \bibinfo {author} {\bibfnamefont {D.}~\bibnamefont {Volfson}}, \ and\
  \bibinfo {author} {\bibfnamefont {L.~S.}\ \bibnamefont {Tsimring}},\ }\href
  {\doibase 10.1103/PhysRevLett.100.058001} {\bibfield  {journal} {\bibinfo
  {journal} {Phys. Rev. Lett.}\ }\textbf {\bibinfo {volume} {100}},\ \bibinfo
  {pages} {058001} (\bibinfo {year} {2008})}\BibitemShut {NoStop}%
\bibitem [{\citenamefont {Deseigne}\ \emph {et~al.}(2010)\citenamefont
  {Deseigne}, \citenamefont {Dauchot},\ and\ \citenamefont
  {Chat\'{e}}}]{Deseigne2010}%
  \BibitemOpen
  \bibfield  {author} {\bibinfo {author} {\bibfnamefont {J.}~\bibnamefont
  {Deseigne}}, \bibinfo {author} {\bibfnamefont {O.}~\bibnamefont {Dauchot}}, \
  and\ \bibinfo {author} {\bibfnamefont {H.}~\bibnamefont {Chat\'{e}}},\ }\href
  {\doibase 10.1103/PhysRevLett.105.098001} {\bibfield  {journal} {\bibinfo
  {journal} {Phys. Rev. Lett.}\ }\textbf {\bibinfo {volume} {105}},\ \bibinfo
  {pages} {098001} (\bibinfo {year} {2010})}\BibitemShut {NoStop}%
\bibitem [{\citenamefont {Kumar}\ \emph {et~al.}(2014)\citenamefont {Kumar},
  \citenamefont {Soni}, \citenamefont {Ramaswamy},\ and\ \citenamefont
  {Sood}}]{Kumar2014}%
  \BibitemOpen
  \bibfield  {author} {\bibinfo {author} {\bibfnamefont {N.}~\bibnamefont
  {Kumar}}, \bibinfo {author} {\bibfnamefont {H.}~\bibnamefont {Soni}},
  \bibinfo {author} {\bibfnamefont {S.}~\bibnamefont {Ramaswamy}}, \ and\
  \bibinfo {author} {\bibfnamefont {A.~K.}\ \bibnamefont {Sood}},\ }\href
  {\doibase 10.1038/ncomms5688} {\bibfield  {journal} {\bibinfo  {journal} {Nat
  Commun}\ }\textbf {\bibinfo {volume} {5}} (\bibinfo {year} {2014}),\
  10.1038/ncomms5688}\BibitemShut {NoStop}%
\bibitem [{\citenamefont {Marchetti}\ \emph {et~al.}(2013)\citenamefont
  {Marchetti}, \citenamefont {Joanny}, \citenamefont {Ramaswamy}, \citenamefont
  {Liverpool}, \citenamefont {Prost}, \citenamefont {Rao},\ and\ \citenamefont
  {Simha}}]{Marchetti2013}%
  \BibitemOpen
  \bibfield  {author} {\bibinfo {author} {\bibfnamefont {M.~C.}\ \bibnamefont
  {Marchetti}}, \bibinfo {author} {\bibfnamefont {J.~F.}\ \bibnamefont
  {Joanny}}, \bibinfo {author} {\bibfnamefont {S.}~\bibnamefont {Ramaswamy}},
  \bibinfo {author} {\bibfnamefont {T.~B.}\ \bibnamefont {Liverpool}}, \bibinfo
  {author} {\bibfnamefont {J.}~\bibnamefont {Prost}}, \bibinfo {author}
  {\bibfnamefont {M.}~\bibnamefont {Rao}}, \ and\ \bibinfo {author}
  {\bibfnamefont {R.~A.}\ \bibnamefont {Simha}},\ }\href {\doibase
  10.1103/RevModPhys.85.1143} {\bibfield  {journal} {\bibinfo  {journal}
  {Reviews of Modern Physics}\ }\textbf {\bibinfo {volume} {85}},\ \bibinfo
  {pages} {1143} (\bibinfo {year} {2013})}\BibitemShut {NoStop}%
\bibitem [{\citenamefont {Bechinger}\ \emph {et~al.}(2016)\citenamefont
  {Bechinger}, \citenamefont {Leonardo}, \citenamefont {Lowen}, \citenamefont
  {Reichhardt}, \citenamefont {Volpe},\ and\ \citenamefont
  {Volpe}}]{Bechinger2016}%
  \BibitemOpen
  \bibfield  {author} {\bibinfo {author} {\bibfnamefont {C.}~\bibnamefont
  {Bechinger}}, \bibinfo {author} {\bibfnamefont {R.~D.}\ \bibnamefont
  {Leonardo}}, \bibinfo {author} {\bibfnamefont {H.}~\bibnamefont {Lowen}},
  \bibinfo {author} {\bibfnamefont {C.}~\bibnamefont {Reichhardt}}, \bibinfo
  {author} {\bibfnamefont {G.}~\bibnamefont {Volpe}}, \ and\ \bibinfo {author}
  {\bibfnamefont {G.}~\bibnamefont {Volpe}},\ }\href
  {http://arXiv.org/abs/1602.00081} {\bibfield  {journal} {\bibinfo  {journal}
  {arXiv:1602.00081}\ } (\bibinfo {year} {2016})}\BibitemShut {NoStop}%
\bibitem [{\citenamefont {Wan}\ \emph {et~al.}(2008)\citenamefont {Wan},
  \citenamefont {Olson~Reichhardt}, \citenamefont {Nussinov},\ and\
  \citenamefont {Reichhardt}}]{Wan2008}%
  \BibitemOpen
  \bibfield  {author} {\bibinfo {author} {\bibfnamefont {M.~B.}\ \bibnamefont
  {Wan}}, \bibinfo {author} {\bibfnamefont {C.~J.}\ \bibnamefont
  {Olson~Reichhardt}}, \bibinfo {author} {\bibfnamefont {Z.}~\bibnamefont
  {Nussinov}}, \ and\ \bibinfo {author} {\bibfnamefont {C.}~\bibnamefont
  {Reichhardt}},\ }\href {\doibase 10.1103/PhysRevLett.101.018102} {\bibfield
  {journal} {\bibinfo  {journal} {Phys. Rev. Lett.}\ }\textbf {\bibinfo
  {volume} {101}},\ \bibinfo {pages} {018102} (\bibinfo {year}
  {2008})}\BibitemShut {NoStop}%
\bibitem [{\citenamefont {Tailleur}\ and\ \citenamefont
  {Cates}(2009)}]{Tailleur2009}%
  \BibitemOpen
  \bibfield  {author} {\bibinfo {author} {\bibfnamefont {J.}~\bibnamefont
  {Tailleur}}\ and\ \bibinfo {author} {\bibfnamefont {M.~E.}\ \bibnamefont
  {Cates}},\ }\href {http://stacks.iop.org/0295-5075/86/i=6/a=60002} {\bibfield
   {journal} {\bibinfo  {journal} {Europhys. Lett.}\ }\textbf {\bibinfo
  {volume} {86}},\ \bibinfo {pages} {60002} (\bibinfo {year}
  {2009})}\BibitemShut {NoStop}%
\bibitem [{\citenamefont {Angelani}\ \emph {et~al.}(2011)\citenamefont
  {Angelani}, \citenamefont {Costanzo},\ and\ \citenamefont
  {Leonardo}}]{Angelani2011}%
  \BibitemOpen
  \bibfield  {author} {\bibinfo {author} {\bibfnamefont {L.}~\bibnamefont
  {Angelani}}, \bibinfo {author} {\bibfnamefont {A.}~\bibnamefont {Costanzo}},
  \ and\ \bibinfo {author} {\bibfnamefont {R.~D.}\ \bibnamefont {Leonardo}},\
  }\href {\doibase 10.1209/0295-5075/96/68002} {\bibfield  {journal} {\bibinfo
  {journal} {Europhys. Lett.}\ }\textbf {\bibinfo {volume} {96}},\ \bibinfo
  {pages} {68002} (\bibinfo {year} {2011})}\BibitemShut {NoStop}%
\bibitem [{\citenamefont {Ghosh}\ \emph {et~al.}(2013)\citenamefont {Ghosh},
  \citenamefont {Misko}, \citenamefont {Marchesoni},\ and\ \citenamefont
  {Nori}}]{Ghosh2013}%
  \BibitemOpen
  \bibfield  {author} {\bibinfo {author} {\bibfnamefont {P.~K.}\ \bibnamefont
  {Ghosh}}, \bibinfo {author} {\bibfnamefont {V.~R.}\ \bibnamefont {Misko}},
  \bibinfo {author} {\bibfnamefont {F.}~\bibnamefont {Marchesoni}}, \ and\
  \bibinfo {author} {\bibfnamefont {F.}~\bibnamefont {Nori}},\ }\href {\doibase
  10.1103/PhysRevLett.110.268301} {\bibfield  {journal} {\bibinfo  {journal}
  {Phys. Rev. Lett.}\ }\textbf {\bibinfo {volume} {110}},\ \bibinfo {pages}
  {268301} (\bibinfo {year} {2013})}\BibitemShut {NoStop}%
\bibitem [{\citenamefont {Ai}\ \emph {et~al.}(2013)\citenamefont {Ai},
  \citenamefont {Chen}, \citenamefont {He}, \citenamefont {Li},\ and\
  \citenamefont {Zhong}}]{Ai2013}%
  \BibitemOpen
  \bibfield  {author} {\bibinfo {author} {\bibfnamefont {B.-q.}\ \bibnamefont
  {Ai}}, \bibinfo {author} {\bibfnamefont {Q.-y.}\ \bibnamefont {Chen}},
  \bibinfo {author} {\bibfnamefont {Y.-f.}\ \bibnamefont {He}}, \bibinfo
  {author} {\bibfnamefont {F.-g.}\ \bibnamefont {Li}}, \ and\ \bibinfo {author}
  {\bibfnamefont {W.-r.}\ \bibnamefont {Zhong}},\ }\href {\doibase
  10.1103/PhysRevE.88.062129} {\bibfield  {journal} {\bibinfo  {journal} {Phys.
  Rev. E}\ }\textbf {\bibinfo {volume} {88}},\ \bibinfo {pages} {062129}
  (\bibinfo {year} {2013})}\BibitemShut {NoStop}%
\bibitem [{\citenamefont {Giomi}\ \emph {et~al.}(2011)\citenamefont {Giomi},
  \citenamefont {Mahadevan}, \citenamefont {Chakraborty},\ and\ \citenamefont
  {Hagan}}]{Giomi2011}%
  \BibitemOpen
  \bibfield  {author} {\bibinfo {author} {\bibfnamefont {L.}~\bibnamefont
  {Giomi}}, \bibinfo {author} {\bibfnamefont {L.}~\bibnamefont {Mahadevan}},
  \bibinfo {author} {\bibfnamefont {B.}~\bibnamefont {Chakraborty}}, \ and\
  \bibinfo {author} {\bibfnamefont {M.~F.}\ \bibnamefont {Hagan}},\ }\href
  {\doibase 10.1103/PhysRevLett.106.218101} {\bibfield  {journal} {\bibinfo
  {journal} {Phys. Rev. Lett.}\ }\textbf {\bibinfo {volume} {106}},\ \bibinfo
  {pages} {218101} (\bibinfo {year} {2011})}\BibitemShut {NoStop}%
\bibitem [{\citenamefont {Giomi}\ and\ \citenamefont
  {Marchetti}(2012)}]{Giomi2012}%
  \BibitemOpen
  \bibfield  {author} {\bibinfo {author} {\bibfnamefont {L.}~\bibnamefont
  {Giomi}}\ and\ \bibinfo {author} {\bibfnamefont {M.~C.}\ \bibnamefont
  {Marchetti}},\ }\href {\doibase 10.1039/C1SM06077E} {\bibfield  {journal}
  {\bibinfo  {journal} {Soft Matter}\ }\textbf {\bibinfo {volume} {8}},\
  \bibinfo {pages} {129} (\bibinfo {year} {2012})}\BibitemShut {NoStop}%
\bibitem [{\citenamefont {Giomi}\ \emph {et~al.}(2012)\citenamefont {Giomi},
  \citenamefont {Mahadevan}, \citenamefont {Chakraborty},\ and\ \citenamefont
  {Hagan}}]{Giomi2012a}%
  \BibitemOpen
  \bibfield  {author} {\bibinfo {author} {\bibfnamefont {L.}~\bibnamefont
  {Giomi}}, \bibinfo {author} {\bibfnamefont {L.}~\bibnamefont {Mahadevan}},
  \bibinfo {author} {\bibfnamefont {B.}~\bibnamefont {Chakraborty}}, \ and\
  \bibinfo {author} {\bibfnamefont {M.~F.}\ \bibnamefont {Hagan}},\ }\href
  {http://stacks.iop.org/0951-7715/25/i=8/a=2245} {\bibfield  {journal}
  {\bibinfo  {journal} {Nonlinearity}\ }\textbf {\bibinfo {volume} {25}},\
  \bibinfo {pages} {2245} (\bibinfo {year} {2012})}\BibitemShut {NoStop}%
\bibitem [{\citenamefont {Voituriez}\ \emph {et~al.}(2006)\citenamefont
  {Voituriez}, \citenamefont {Joanny},\ and\ \citenamefont
  {Prost}}]{Voituriez2006}%
  \BibitemOpen
  \bibfield  {author} {\bibinfo {author} {\bibfnamefont {R.}~\bibnamefont
  {Voituriez}}, \bibinfo {author} {\bibfnamefont {J.}~\bibnamefont {Joanny}}, \
  and\ \bibinfo {author} {\bibfnamefont {J.}~\bibnamefont {Prost}},\ }\href
  {\doibase 10.1103/PhysRevLett.96.028102} {\bibfield  {journal} {\bibinfo
  {journal} {Phys. Rev. Lett.}\ }\textbf {\bibinfo {volume} {96}},\ \bibinfo
  {pages} {028102} (\bibinfo {year} {2006})}\BibitemShut {NoStop}%
\bibitem [{\citenamefont {Marenduzzo}\ \emph {et~al.}(2007)\citenamefont
  {Marenduzzo}, \citenamefont {Orlandini},\ and\ \citenamefont
  {Yeomans}}]{Marenduzzo2007a}%
  \BibitemOpen
  \bibfield  {author} {\bibinfo {author} {\bibfnamefont {D.}~\bibnamefont
  {Marenduzzo}}, \bibinfo {author} {\bibfnamefont {E.}~\bibnamefont
  {Orlandini}}, \ and\ \bibinfo {author} {\bibfnamefont {J.}~\bibnamefont
  {Yeomans}},\ }\href {\doibase 10.1103/PhysRevLett.98.118102} {\bibfield
  {journal} {\bibinfo  {journal} {Phys. Rev. Lett.}\ }\textbf {\bibinfo
  {volume} {98}},\ \bibinfo {pages} {118102} (\bibinfo {year}
  {2007})}\BibitemShut {NoStop}%
\bibitem [{\citenamefont {Chat\'{e}}\ \emph {et~al.}(2006)\citenamefont
  {Chat\'{e}}, \citenamefont {Ginelli},\ and\ \citenamefont
  {Montagne}}]{Ginelli2006}%
  \BibitemOpen
  \bibfield  {author} {\bibinfo {author} {\bibfnamefont {H.}~\bibnamefont
  {Chat\'{e}}}, \bibinfo {author} {\bibfnamefont {F.}~\bibnamefont {Ginelli}},
  \ and\ \bibinfo {author} {\bibfnamefont {R.}~\bibnamefont {Montagne}},\
  }\href {\doibase 10.1103/PhysRevLett.96.180602} {\bibfield  {journal}
  {\bibinfo  {journal} {Phys. Rev. Lett.}\ }\textbf {\bibinfo {volume} {96}},\
  \bibinfo {pages} {180602} (\bibinfo {year} {2006})}\BibitemShut {NoStop}%
\bibitem [{\citenamefont {Fielding}\ \emph {et~al.}(2011)\citenamefont
  {Fielding}, \citenamefont {Marenduzzo},\ and\ \citenamefont
  {Cates}}]{Fielding2011}%
  \BibitemOpen
  \bibfield  {author} {\bibinfo {author} {\bibfnamefont {S.~M.}\ \bibnamefont
  {Fielding}}, \bibinfo {author} {\bibfnamefont {D.}~\bibnamefont
  {Marenduzzo}}, \ and\ \bibinfo {author} {\bibfnamefont {M.~E.}\ \bibnamefont
  {Cates}},\ }\href {\doibase 10.1103/PhysRevE.83.041910} {\bibfield  {journal}
  {\bibinfo  {journal} {Phys. Rev. E}\ }\textbf {\bibinfo {volume} {83}},\
  \bibinfo {pages} {041910} (\bibinfo {year} {2011})}\BibitemShut {NoStop}%
\bibitem [{\citenamefont {Marenduzzo}\ \emph {et~al.}(2008)\citenamefont
  {Marenduzzo}, \citenamefont {Orlandini}, \citenamefont {Cates},\ and\
  \citenamefont {Yeomans}}]{Marenduzzo2008}%
  \BibitemOpen
  \bibfield  {author} {\bibinfo {author} {\bibfnamefont {D.}~\bibnamefont
  {Marenduzzo}}, \bibinfo {author} {\bibfnamefont {E.}~\bibnamefont
  {Orlandini}}, \bibinfo {author} {\bibfnamefont {M.}~\bibnamefont {Cates}}, \
  and\ \bibinfo {author} {\bibfnamefont {J.}~\bibnamefont {Yeomans}},\ }\href
  {\doibase 10.1016/j.jnnfm.2007.02.005} {\bibfield  {journal} {\bibinfo
  {journal} {J. Non-Newtonian Fluid Mech.}\ }\textbf {\bibinfo {volume}
  {149}},\ \bibinfo {pages} {56} (\bibinfo {year} {2008})}\BibitemShut
  {NoStop}%
\bibitem [{\citenamefont {Giomi}\ \emph {et~al.}(2008)\citenamefont {Giomi},
  \citenamefont {Marchetti},\ and\ \citenamefont {Liverpool}}]{Giomi2008}%
  \BibitemOpen
  \bibfield  {author} {\bibinfo {author} {\bibfnamefont {L.}~\bibnamefont
  {Giomi}}, \bibinfo {author} {\bibfnamefont {M.}~\bibnamefont {Marchetti}}, \
  and\ \bibinfo {author} {\bibfnamefont {T.}~\bibnamefont {Liverpool}},\ }\href
  {\doibase 10.1103/PhysRevLett.101.198101} {\bibfield  {journal} {\bibinfo
  {journal} {Phys. Rev. Lett.}\ }\textbf {\bibinfo {volume} {101}},\ \bibinfo
  {pages} {198101} (\bibinfo {year} {2008})}\BibitemShut {NoStop}%
\bibitem [{\citenamefont {Fily}\ and\ \citenamefont
  {Marchetti}(2012)}]{Fily2012}%
  \BibitemOpen
  \bibfield  {author} {\bibinfo {author} {\bibfnamefont {Y.}~\bibnamefont
  {Fily}}\ and\ \bibinfo {author} {\bibfnamefont {M.~C.}\ \bibnamefont
  {Marchetti}},\ }\href {\doibase 10.1103/PhysRevLett.108.235702} {\bibfield
  {journal} {\bibinfo  {journal} {Phys. Rev. Lett.}\ }\textbf {\bibinfo
  {volume} {108}},\ \bibinfo {pages} {235702} (\bibinfo {year}
  {2012})}\BibitemShut {NoStop}%
\bibitem [{\citenamefont {Redner}\ \emph
  {et~al.}(2013{\natexlab{a}})\citenamefont {Redner}, \citenamefont {Hagan},\
  and\ \citenamefont {Baskaran}}]{Redner2013}%
  \BibitemOpen
  \bibfield  {author} {\bibinfo {author} {\bibfnamefont {G.~S.}\ \bibnamefont
  {Redner}}, \bibinfo {author} {\bibfnamefont {M.~F.}\ \bibnamefont {Hagan}}, \
  and\ \bibinfo {author} {\bibfnamefont {A.}~\bibnamefont {Baskaran}},\ }\href
  {\doibase 10.1103/PhysRevLett.110.055701} {\bibfield  {journal} {\bibinfo
  {journal} {Phys. Rev. Lett.}\ }\textbf {\bibinfo {volume} {110}},\ \bibinfo
  {pages} {055701} (\bibinfo {year} {2013}{\natexlab{a}})}\BibitemShut
  {NoStop}%
\bibitem [{\citenamefont {Stenhammar}\ \emph {et~al.}(2013)\citenamefont
  {Stenhammar}, \citenamefont {Tiribocchi}, \citenamefont {Allen},
  \citenamefont {Marenduzzo},\ and\ \citenamefont {Cates}}]{Stenhammar2013}%
  \BibitemOpen
  \bibfield  {author} {\bibinfo {author} {\bibfnamefont {J.}~\bibnamefont
  {Stenhammar}}, \bibinfo {author} {\bibfnamefont {A.}~\bibnamefont
  {Tiribocchi}}, \bibinfo {author} {\bibfnamefont {R.~J.}\ \bibnamefont
  {Allen}}, \bibinfo {author} {\bibfnamefont {D.}~\bibnamefont {Marenduzzo}}, \
  and\ \bibinfo {author} {\bibfnamefont {M.~E.}\ \bibnamefont {Cates}},\ }\href
  {\doibase 10.1103/PhysRevLett.111.145702} {\bibfield  {journal} {\bibinfo
  {journal} {Phys. Rev. Lett.}\ }\textbf {\bibinfo {volume} {111}},\ \bibinfo
  {pages} {145702} (\bibinfo {year} {2013})}\BibitemShut {NoStop}%
\bibitem [{\citenamefont {Buttinoni}\ \emph {et~al.}(2013)\citenamefont
  {Buttinoni}, \citenamefont {Bialk\'{e}}, \citenamefont {K\"{u}mmel},
  \citenamefont {L\"{o}wen}, \citenamefont {Bechinger},\ and\ \citenamefont
  {Speck}}]{Buttinoni2013a}%
  \BibitemOpen
  \bibfield  {author} {\bibinfo {author} {\bibfnamefont {I.}~\bibnamefont
  {Buttinoni}}, \bibinfo {author} {\bibfnamefont {J.}~\bibnamefont
  {Bialk\'{e}}}, \bibinfo {author} {\bibfnamefont {F.}~\bibnamefont
  {K\"{u}mmel}}, \bibinfo {author} {\bibfnamefont {H.}~\bibnamefont
  {L\"{o}wen}}, \bibinfo {author} {\bibfnamefont {C.}~\bibnamefont
  {Bechinger}}, \ and\ \bibinfo {author} {\bibfnamefont {T.}~\bibnamefont
  {Speck}},\ }\href {\doibase 10.1103/PhysRevLett.110.238301} {\bibfield
  {journal} {\bibinfo  {journal} {Phys. Rev. Lett.}\ }\textbf {\bibinfo
  {volume} {110}},\ \bibinfo {pages} {238301} (\bibinfo {year}
  {2013})}\BibitemShut {NoStop}%
\bibitem [{\citenamefont {Mognetti}\ \emph {et~al.}(2013)\citenamefont
  {Mognetti}, \citenamefont {A.}, \citenamefont {Angioletti-Uberti},
  \citenamefont {Cacciuto}, \citenamefont {Valeriani},\ and\ \citenamefont
  {Frenkel}}]{Mognetti2013}%
  \BibitemOpen
  \bibfield  {author} {\bibinfo {author} {\bibfnamefont {B.~M.}\ \bibnamefont
  {Mognetti}}, \bibinfo {author} {\bibfnamefont {S.}~\bibnamefont {A.}},
  \bibinfo {author} {\bibfnamefont {S.}~\bibnamefont {Angioletti-Uberti}},
  \bibinfo {author} {\bibfnamefont {A.}~\bibnamefont {Cacciuto}}, \bibinfo
  {author} {\bibfnamefont {C.}~\bibnamefont {Valeriani}}, \ and\ \bibinfo
  {author} {\bibfnamefont {D.}~\bibnamefont {Frenkel}},\ }\href {\doibase
  10.1103/PhysRevLett.111.245702} {\bibfield  {journal} {\bibinfo  {journal}
  {Phys. Rev. Lett.}\ }\textbf {\bibinfo {volume} {111}},\ \bibinfo {pages}
  {245702} (\bibinfo {year} {2013})}\BibitemShut {NoStop}%
\bibitem [{\citenamefont {Stenhammar}\ \emph {et~al.}(2014)\citenamefont
  {Stenhammar}, \citenamefont {Marenduzzo}, \citenamefont {Allen},\ and\
  \citenamefont {Cates}}]{Stenhammar2014b}%
  \BibitemOpen
  \bibfield  {author} {\bibinfo {author} {\bibfnamefont {J.}~\bibnamefont
  {Stenhammar}}, \bibinfo {author} {\bibfnamefont {D.}~\bibnamefont
  {Marenduzzo}}, \bibinfo {author} {\bibfnamefont {R.~J.}\ \bibnamefont
  {Allen}}, \ and\ \bibinfo {author} {\bibfnamefont {M.~E.}\ \bibnamefont
  {Cates}},\ }\href {\doibase 10.1039/c3sm52813h} {\bibfield  {journal}
  {\bibinfo  {journal} {Soft Matter}\ }\textbf {\bibinfo {volume} {10}},\
  \bibinfo {pages} {1489} (\bibinfo {year} {2014})}\BibitemShut {NoStop}%
\bibitem [{\citenamefont {Wysocki}\ \emph {et~al.}(2014)\citenamefont
  {Wysocki}, \citenamefont {Winkler},\ and\ \citenamefont
  {Gompper}}]{Wysocki2014}%
  \BibitemOpen
  \bibfield  {author} {\bibinfo {author} {\bibfnamefont {A.}~\bibnamefont
  {Wysocki}}, \bibinfo {author} {\bibfnamefont {R.~G.}\ \bibnamefont
  {Winkler}}, \ and\ \bibinfo {author} {\bibfnamefont {G.}~\bibnamefont
  {Gompper}},\ }\href {http://stacks.iop.org/0295-5075/105/i=4/a=48004}
  {\bibfield  {journal} {\bibinfo  {journal} {Europhys. Lett.}\ }\textbf
  {\bibinfo {volume} {105}},\ \bibinfo {pages} {48004} (\bibinfo {year}
  {2014})}\BibitemShut {NoStop}%
\bibitem [{\citenamefont {Cates}\ and\ \citenamefont
  {Tailleur}(2015)}]{Cates2015}%
  \BibitemOpen
  \bibfield  {author} {\bibinfo {author} {\bibfnamefont {M.~E.}\ \bibnamefont
  {Cates}}\ and\ \bibinfo {author} {\bibfnamefont {J.}~\bibnamefont
  {Tailleur}},\ }\href {\doibase 10.1146/annurev-conmatphys-031214-014710}
  {\bibfield  {journal} {\bibinfo  {journal} {Annu. Rev. Condens. Matter
  Phys.}\ }\textbf {\bibinfo {volume} {6}},\ \bibinfo {pages} {219} (\bibinfo
  {year} {2015})},\ \Eprint
  {http://arxiv.org/abs/http://dx.doi.org/10.1146/annurev-conmatphys-031214-014710}
  {http://dx.doi.org/10.1146/annurev-conmatphys-031214-014710} \BibitemShut
  {NoStop}%
\bibitem [{\citenamefont {Marchetti}\ \emph {et~al.}(2015)\citenamefont
  {Marchetti}, \citenamefont {Fily}, \citenamefont {Henkes}, \citenamefont
  {Patch},\ and\ \citenamefont {Yllanes}}]{marchetti2015structure}%
  \BibitemOpen
  \bibfield  {author} {\bibinfo {author} {\bibfnamefont {M.~C.}\ \bibnamefont
  {Marchetti}}, \bibinfo {author} {\bibfnamefont {Y.}~\bibnamefont {Fily}},
  \bibinfo {author} {\bibfnamefont {S.}~\bibnamefont {Henkes}}, \bibinfo
  {author} {\bibfnamefont {A.}~\bibnamefont {Patch}}, \ and\ \bibinfo {author}
  {\bibfnamefont {D.}~\bibnamefont {Yllanes}},\ }\href@noop {} {\bibfield
  {journal} {\bibinfo  {journal} {arXiv preprint arXiv:1510.00425}\ } (\bibinfo
  {year} {2015})}\BibitemShut {NoStop}%
\bibitem [{\citenamefont {Stenhammar}\ \emph {et~al.}(2015)\citenamefont
  {Stenhammar}, \citenamefont {Wittkowski}, \citenamefont {Marenduzzo},\ and\
  \citenamefont {Cates}}]{Stenhammar2015}%
  \BibitemOpen
  \bibfield  {author} {\bibinfo {author} {\bibfnamefont {J.}~\bibnamefont
  {Stenhammar}}, \bibinfo {author} {\bibfnamefont {R.}~\bibnamefont
  {Wittkowski}}, \bibinfo {author} {\bibfnamefont {D.}~\bibnamefont
  {Marenduzzo}}, \ and\ \bibinfo {author} {\bibfnamefont {M.~E.}\ \bibnamefont
  {Cates}},\ }\href {\doibase 10.1103/PhysRevLett.114.018301} {\bibfield
  {journal} {\bibinfo  {journal} {Phys. Rev. Lett.}\ }\textbf {\bibinfo
  {volume} {114}},\ \bibinfo {pages} {018301} (\bibinfo {year}
  {2015})}\BibitemShut {NoStop}%
\bibitem [{\citenamefont {Ni}\ \emph {et~al.}(2013)\citenamefont {Ni},
  \citenamefont {Stuart}, \citenamefont {Dijkstra},\ and\ \citenamefont
  {Bolhuis}}]{Ni2013a}%
  \BibitemOpen
  \bibfield  {author} {\bibinfo {author} {\bibfnamefont {R.}~\bibnamefont
  {Ni}}, \bibinfo {author} {\bibfnamefont {M.~a.~C.}\ \bibnamefont {Stuart}},
  \bibinfo {author} {\bibfnamefont {M.}~\bibnamefont {Dijkstra}}, \ and\
  \bibinfo {author} {\bibfnamefont {P.~G.}\ \bibnamefont {Bolhuis}},\ }\href
  {\doibase 10.1039/C4SM01015A} {\bibfield  {journal} {\bibinfo  {journal}
  {Soft Matter}\ }\textbf {\bibinfo {volume} {10}},\ \bibinfo {pages} {6609}
  (\bibinfo {year} {2013})}\BibitemShut {NoStop}%
\bibitem [{\citenamefont {Solon}\ \emph
  {et~al.}(2015{\natexlab{a}})\citenamefont {Solon}, \citenamefont
  {Stenhammar}, \citenamefont {Wittkowski}, \citenamefont {Kardar},
  \citenamefont {Kafri}, \citenamefont {Cates},\ and\ \citenamefont
  {Tailleur}}]{Solon2015}%
  \BibitemOpen
  \bibfield  {author} {\bibinfo {author} {\bibfnamefont {A.~P.}\ \bibnamefont
  {Solon}}, \bibinfo {author} {\bibfnamefont {J.}~\bibnamefont {Stenhammar}},
  \bibinfo {author} {\bibfnamefont {R.}~\bibnamefont {Wittkowski}}, \bibinfo
  {author} {\bibfnamefont {M.}~\bibnamefont {Kardar}}, \bibinfo {author}
  {\bibfnamefont {Y.}~\bibnamefont {Kafri}}, \bibinfo {author} {\bibfnamefont
  {M.~E.}\ \bibnamefont {Cates}}, \ and\ \bibinfo {author} {\bibfnamefont
  {J.}~\bibnamefont {Tailleur}},\ }\href {\doibase
  10.1103/PhysRevLett.114.198301} {\bibfield  {journal} {\bibinfo  {journal}
  {Phys. Rev. Lett.}\ }\textbf {\bibinfo {volume} {114}},\ \bibinfo {pages}
  {198301} (\bibinfo {year} {2015}{\natexlab{a}})}\BibitemShut {NoStop}%
\bibitem [{\citenamefont {Solon}\ \emph
  {et~al.}(2015{\natexlab{b}})\citenamefont {Solon}, \citenamefont {Fily},
  \citenamefont {Baskaran}, \citenamefont {Cates}, \citenamefont {Kafri},
  \citenamefont {Kardar},\ and\ \citenamefont {Tailleur}}]{Solon2015b}%
  \BibitemOpen
  \bibfield  {author} {\bibinfo {author} {\bibfnamefont {a.~P.}\ \bibnamefont
  {Solon}}, \bibinfo {author} {\bibfnamefont {Y.}~\bibnamefont {Fily}},
  \bibinfo {author} {\bibfnamefont {A.}~\bibnamefont {Baskaran}}, \bibinfo
  {author} {\bibfnamefont {M.~E.}\ \bibnamefont {Cates}}, \bibinfo {author}
  {\bibfnamefont {Y.}~\bibnamefont {Kafri}}, \bibinfo {author} {\bibfnamefont
  {M.}~\bibnamefont {Kardar}}, \ and\ \bibinfo {author} {\bibfnamefont
  {J.}~\bibnamefont {Tailleur}},\ }\href {\doibase 10.1038/nphys3377}
  {\bibfield  {journal} {\bibinfo  {journal} {Nat. Phys.}\ ,\ \bibinfo {pages}
  {4}} (\bibinfo {year} {2015}{\natexlab{b}})}\BibitemShut {NoStop}%
\bibitem [{\citenamefont {Takatori}\ \emph {et~al.}(2014)\citenamefont
  {Takatori}, \citenamefont {Yan},\ and\ \citenamefont {Brady}}]{Takatori2014}%
  \BibitemOpen
  \bibfield  {author} {\bibinfo {author} {\bibfnamefont {S.~C.}\ \bibnamefont
  {Takatori}}, \bibinfo {author} {\bibfnamefont {W.}~\bibnamefont {Yan}}, \
  and\ \bibinfo {author} {\bibfnamefont {J.~F.}\ \bibnamefont {Brady}},\ }\href
  {\doibase 10.1103/PhysRevLett.113.028103} {\bibfield  {journal} {\bibinfo
  {journal} {Phys. Rev. Lett.}\ }\textbf {\bibinfo {volume} {113}},\ \bibinfo
  {pages} {028103} (\bibinfo {year} {2014})}\BibitemShut {NoStop}%
\bibitem [{\citenamefont {Takatori}\ and\ \citenamefont
  {Brady}(2015)}]{Takatori2015}%
  \BibitemOpen
  \bibfield  {author} {\bibinfo {author} {\bibfnamefont {S.~C.}\ \bibnamefont
  {Takatori}}\ and\ \bibinfo {author} {\bibfnamefont {J.~F.}\ \bibnamefont
  {Brady}},\ }\href@noop {} {\bibfield  {journal} {\bibinfo  {journal}
  {Physical Review E}\ }\textbf {\bibinfo {volume} {91}},\ \bibinfo {pages}
  {032117} (\bibinfo {year} {2015})}\BibitemShut {NoStop}%
\bibitem [{\citenamefont {Speck}\ and\ \citenamefont {Jack}(2015)}]{Speck2015}%
  \BibitemOpen
  \bibfield  {author} {\bibinfo {author} {\bibfnamefont {T.}~\bibnamefont
  {Speck}}\ and\ \bibinfo {author} {\bibfnamefont {R.~L.}\ \bibnamefont
  {Jack}},\ }\href@noop {} {\bibfield  {journal} {\bibinfo  {journal} {arXiv
  preprint arXiv:1512.00830}\ } (\bibinfo {year} {2015})}\BibitemShut {NoStop}%
\bibitem [{\citenamefont {Speck}\ \emph {et~al.}(2015)\citenamefont {Speck},
  \citenamefont {Menzel}, \citenamefont {Bialk\'{e}},\ and\ \citenamefont
  {L\"{o}wen}}]{Speck2015a}%
  \BibitemOpen
  \bibfield  {author} {\bibinfo {author} {\bibfnamefont {T.}~\bibnamefont
  {Speck}}, \bibinfo {author} {\bibfnamefont {A.~M.}\ \bibnamefont {Menzel}},
  \bibinfo {author} {\bibfnamefont {J.}~\bibnamefont {Bialk\'{e}}}, \ and\
  \bibinfo {author} {\bibfnamefont {H.}~\bibnamefont {L\"{o}wen}},\ }\href
  {\doibase 10.1063/1.4922324} {\bibfield  {journal} {\bibinfo  {journal} {J.
  Chem. Phys.}\ }\textbf {\bibinfo {volume} {142}},\ \bibinfo {pages} {224109}
  (\bibinfo {year} {2015})}\BibitemShut {NoStop}%
\bibitem [{\citenamefont {Winkler}\ \emph {et~al.}(2015)\citenamefont
  {Winkler}, \citenamefont {Wysocki},\ and\ \citenamefont
  {Gompper}}]{Winkler2015}%
  \BibitemOpen
  \bibfield  {author} {\bibinfo {author} {\bibfnamefont {R.~G.}\ \bibnamefont
  {Winkler}}, \bibinfo {author} {\bibfnamefont {A.}~\bibnamefont {Wysocki}}, \
  and\ \bibinfo {author} {\bibfnamefont {G.}~\bibnamefont {Gompper}},\
  }\href@noop {} {\  (\bibinfo {year} {2015})},\ \Eprint
  {http://arxiv.org/abs/arXiv:1506.03941} {arXiv:arXiv:1506.03941} \BibitemShut
  {NoStop}%
\bibitem [{\citenamefont {Ginot}\ \emph {et~al.}(2015)\citenamefont {Ginot},
  \citenamefont {Theurkauff}, \citenamefont {Levis}, \citenamefont {Ybert},
  \citenamefont {Bocquet}, \citenamefont {Berthier}, \citenamefont
  {Cottin-bizonne}, \citenamefont {Mati\`{e}re}, \citenamefont {Umr},
  \citenamefont {Claude}, \citenamefont {Lyon}, \citenamefont {Lyon},\ and\
  \citenamefont {Cedex}}]{Ginot2015}%
  \BibitemOpen
  \bibfield  {author} {\bibinfo {author} {\bibfnamefont {F.}~\bibnamefont
  {Ginot}}, \bibinfo {author} {\bibfnamefont {I.}~\bibnamefont {Theurkauff}},
  \bibinfo {author} {\bibfnamefont {D.}~\bibnamefont {Levis}}, \bibinfo
  {author} {\bibfnamefont {C.}~\bibnamefont {Ybert}}, \bibinfo {author}
  {\bibfnamefont {L.}~\bibnamefont {Bocquet}}, \bibinfo {author} {\bibfnamefont
  {L.}~\bibnamefont {Berthier}}, \bibinfo {author} {\bibfnamefont
  {C.}~\bibnamefont {Cottin-bizonne}}, \bibinfo {author} {\bibfnamefont
  {I.~L.}\ \bibnamefont {Mati\`{e}re}}, \bibinfo {author} {\bibfnamefont
  {C.}~\bibnamefont {Umr}}, \bibinfo {author} {\bibfnamefont {U.}~\bibnamefont
  {Claude}}, \bibinfo {author} {\bibfnamefont {B.}~\bibnamefont {Lyon}},
  \bibinfo {author} {\bibfnamefont {U.~D.}\ \bibnamefont {Lyon}}, \ and\
  \bibinfo {author} {\bibfnamefont {V.}~\bibnamefont {Cedex}},\ }\href
  {\doibase 10.1103/PhysRevX.5.011004} {\bibfield  {journal} {\bibinfo
  {journal} {Phys. Rev. X}\ }\textbf {\bibinfo {volume} {5}},\ \bibinfo {pages}
  {011004} (\bibinfo {year} {2015})}\BibitemShut {NoStop}%
\bibitem [{\citenamefont {Bialk\'{e}}\ \emph {et~al.}(2014)\citenamefont
  {Bialk\'{e}}, \citenamefont {Speck},\ and\ \citenamefont
  {L\"{o}wen}}]{Bialke2014}%
  \BibitemOpen
  \bibfield  {author} {\bibinfo {author} {\bibfnamefont {J.}~\bibnamefont
  {Bialk\'{e}}}, \bibinfo {author} {\bibfnamefont {T.}~\bibnamefont {Speck}}, \
  and\ \bibinfo {author} {\bibfnamefont {H.}~\bibnamefont {L\"{o}wen}},\ }\href
  {\doibase 10.1016/j.jnoncrysol.2014.08.011} {\bibfield  {journal} {\bibinfo
  {journal} {J. Non-Cryst. Solids}\ }\textbf {\bibinfo {volume} {407}},\
  \bibinfo {pages} {11} (\bibinfo {year} {2014})}\BibitemShut {NoStop}%
\bibitem [{\citenamefont {Wysocki}\ \emph {et~al.}(2013)\citenamefont
  {Wysocki}, \citenamefont {Winkler},\ and\ \citenamefont
  {Gompper}}]{Wysocki2013}%
  \BibitemOpen
  \bibfield  {author} {\bibinfo {author} {\bibfnamefont {A.}~\bibnamefont
  {Wysocki}}, \bibinfo {author} {\bibfnamefont {R.~G.}\ \bibnamefont
  {Winkler}}, \ and\ \bibinfo {author} {\bibfnamefont {G.}~\bibnamefont
  {Gompper}},\ }\href {http://arxiv.org/abs/1308.6423v1} {\enquote {\bibinfo
  {title} {Cooperative motion of active brownian spheres in three-dimensional
  dense suspensions},}\ } (\bibinfo {year} {2013})\BibitemShut {NoStop}%
\bibitem [{\citenamefont {Tailleur}\ and\ \citenamefont
  {Cates}(2008)}]{Tailleur2008}%
  \BibitemOpen
  \bibfield  {author} {\bibinfo {author} {\bibfnamefont {J.}~\bibnamefont
  {Tailleur}}\ and\ \bibinfo {author} {\bibfnamefont {M.}~\bibnamefont
  {Cates}},\ }\href {\doibase 10.1103/PhysRevLett.100.218103} {\bibfield
  {journal} {\bibinfo  {journal} {Phys. Rev. Lett.}\ }\textbf {\bibinfo
  {volume} {100}},\ \bibinfo {pages} {218103} (\bibinfo {year}
  {2008})}\BibitemShut {NoStop}%
\bibitem [{\citenamefont {Wittkowski}\ \emph {et~al.}(2014)\citenamefont
  {Wittkowski}, \citenamefont {Tiribocchi}, \citenamefont {Stenhammar},
  \citenamefont {Allen}, \citenamefont {Marenduzzo},\ and\ \citenamefont
  {Cates}}]{Wittkowski2014}%
  \BibitemOpen
  \bibfield  {author} {\bibinfo {author} {\bibfnamefont {R.}~\bibnamefont
  {Wittkowski}}, \bibinfo {author} {\bibfnamefont {A.}~\bibnamefont
  {Tiribocchi}}, \bibinfo {author} {\bibfnamefont {J.}~\bibnamefont
  {Stenhammar}}, \bibinfo {author} {\bibfnamefont {R.~J.}\ \bibnamefont
  {Allen}}, \bibinfo {author} {\bibfnamefont {D.}~\bibnamefont {Marenduzzo}}, \
  and\ \bibinfo {author} {\bibfnamefont {M.~E.}\ \bibnamefont {Cates}},\ }\href
  {\doibase 10.1038/ncomms5351} {\bibfield  {journal} {\bibinfo  {journal}
  {Nat. Comm.}\ }\textbf {\bibinfo {volume} {5}},\ \bibinfo {pages} {4351}
  (\bibinfo {year} {2014})}\BibitemShut {NoStop}%
\bibitem [{\citenamefont {Speck}\ \emph {et~al.}(2014)\citenamefont {Speck},
  \citenamefont {Bialk\'{e}}, \citenamefont {Menzel},\ and\ \citenamefont
  {L\"{o}wen}}]{Speck2014}%
  \BibitemOpen
  \bibfield  {author} {\bibinfo {author} {\bibfnamefont {T.}~\bibnamefont
  {Speck}}, \bibinfo {author} {\bibfnamefont {J.}~\bibnamefont {Bialk\'{e}}},
  \bibinfo {author} {\bibfnamefont {A.~M.}\ \bibnamefont {Menzel}}, \ and\
  \bibinfo {author} {\bibfnamefont {H.}~\bibnamefont {L\"{o}wen}},\ }\href
  {\doibase 10.1103/PhysRevLett.112.218304} {\bibfield  {journal} {\bibinfo
  {journal} {Phys. Rev. Lett.}\ }\textbf {\bibinfo {volume} {112}},\ \bibinfo
  {pages} {218304} (\bibinfo {year} {2014})}\BibitemShut {NoStop}%
\bibitem [{\citenamefont {Bialk\'{e}}\ \emph {et~al.}(2015)\citenamefont
  {Bialk\'{e}}, \citenamefont {Siebert}, \citenamefont {L\"{o}wen},\ and\
  \citenamefont {Speck}}]{Bialke2015}%
  \BibitemOpen
  \bibfield  {author} {\bibinfo {author} {\bibfnamefont {J.}~\bibnamefont
  {Bialk\'{e}}}, \bibinfo {author} {\bibfnamefont {J.~T.}\ \bibnamefont
  {Siebert}}, \bibinfo {author} {\bibfnamefont {H.}~\bibnamefont {L\"{o}wen}},
  \ and\ \bibinfo {author} {\bibfnamefont {T.}~\bibnamefont {Speck}},\ }\href
  {\doibase 10.1103/PhysRevLett.115.098301} {\bibfield  {journal} {\bibinfo
  {journal} {Phys. Rev. Lett.}\ }\textbf {\bibinfo {volume} {115}},\ \bibinfo
  {pages} {098301} (\bibinfo {year} {2015})}\BibitemShut {NoStop}%
\bibitem [{\citenamefont {Oxtoby}(1992)}]{Oxtoby1992}%
  \BibitemOpen
  \bibfield  {author} {\bibinfo {author} {\bibfnamefont {D.~W.}\ \bibnamefont
  {Oxtoby}},\ }\href {http://stacks.iop.org/0953-8984/4/i=38/a=001} {\bibfield
  {journal} {\bibinfo  {journal} {J. Phys.: Condens. Matter}\ }\textbf
  {\bibinfo {volume} {4}},\ \bibinfo {pages} {7627} (\bibinfo {year}
  {1992})}\BibitemShut {NoStop}%
\bibitem [{\citenamefont {Agarwal}\ and\ \citenamefont
  {Peters}(2014)}]{Agarwal2014}%
  \BibitemOpen
  \bibfield  {author} {\bibinfo {author} {\bibfnamefont {V.}~\bibnamefont
  {Agarwal}}\ and\ \bibinfo {author} {\bibfnamefont {B.}~\bibnamefont
  {Peters}},\ }\enquote {\bibinfo {title} {Solute precipitate nucleation: A
  review of theory and simulation advances},}\ in\ \href {\doibase
  10.1002/9781118755815.ch03} {\emph {\bibinfo {booktitle} {Advances in
  Chemical Physics: Volume 155}}}\ (\bibinfo  {publisher} {John Wiley \& Sons,
  Inc.},\ \bibinfo {year} {2014})\ pp.\ \bibinfo {pages} {97--160}\BibitemShut
  {NoStop}%
\bibitem [{\citenamefont {Yoreo}\ and\ \citenamefont
  {Vekilov}(2003)}]{Yoreo2003}%
  \BibitemOpen
  \bibfield  {author} {\bibinfo {author} {\bibfnamefont {J.~J.~D.}\
  \bibnamefont {Yoreo}}\ and\ \bibinfo {author} {\bibfnamefont {P.~G.}\
  \bibnamefont {Vekilov}},\ }\href@noop {} {\bibfield  {journal} {\bibinfo
  {journal} {Reviews in Mineralogy and Geochemistry}\ }\textbf {\bibinfo
  {volume} {54}},\ \bibinfo {pages} {57} (\bibinfo {year} {2003})}\BibitemShut
  {NoStop}%
\bibitem [{\citenamefont {Becker}\ and\ \citenamefont
  {D{\"o}ring}(1935)}]{Becker1935}%
  \BibitemOpen
  \bibfield  {author} {\bibinfo {author} {\bibfnamefont {R.}~\bibnamefont
  {Becker}}\ and\ \bibinfo {author} {\bibfnamefont {W.}~\bibnamefont
  {D{\"o}ring}},\ }\href@noop {} {\bibfield  {journal} {\bibinfo  {journal}
  {Ann. Phys. (Leipzig)}\ }\textbf {\bibinfo {volume} {416}},\ \bibinfo {pages}
  {719} (\bibinfo {year} {1935})}\BibitemShut {NoStop}%
\bibitem [{\citenamefont {Redner}\ \emph
  {et~al.}(2013{\natexlab{b}})\citenamefont {Redner}, \citenamefont
  {Baskaran},\ and\ \citenamefont {Hagan}}]{Redner2013a}%
  \BibitemOpen
  \bibfield  {author} {\bibinfo {author} {\bibfnamefont {G.~S.}\ \bibnamefont
  {Redner}}, \bibinfo {author} {\bibfnamefont {A.}~\bibnamefont {Baskaran}}, \
  and\ \bibinfo {author} {\bibfnamefont {M.~F.}\ \bibnamefont {Hagan}},\ }\href
  {\doibase 10.1103/PhysRevE.88.012305} {\bibfield  {journal} {\bibinfo
  {journal} {Phys. Rev. E}\ }\textbf {\bibinfo {volume} {88}},\ \bibinfo
  {pages} {012305} (\bibinfo {year} {2013}{\natexlab{b}})}\BibitemShut
  {NoStop}%
\bibitem [{\citenamefont {Richard}\ \emph {et~al.}(2016)\citenamefont
  {Richard}, \citenamefont {Lowen},\ and\ \citenamefont {Speck}}]{Richard2016}%
  \BibitemOpen
  \bibfield  {author} {\bibinfo {author} {\bibfnamefont {D.}~\bibnamefont
  {Richard}}, \bibinfo {author} {\bibfnamefont {H.}~\bibnamefont {Lowen}}, \
  and\ \bibinfo {author} {\bibfnamefont {T.}~\bibnamefont {Speck}},\ }\href
  {\doibase 10.1039/C6SM00485G} {\bibfield  {journal} {\bibinfo  {journal}
  {Soft Matter}\ } (\bibinfo {year} {2016}),\ 10.1039/C6SM00485G}\BibitemShut
  {NoStop}%
\bibitem [{SIr(tion)}]{SIref}%
  \BibitemOpen
  \href@noop {} {} (\bibinfo {year} {Supplementary Information}),\ \bibinfo
  {note} {supplemental figures and additional model details are included at the
  end of this document}\BibitemShut {NoStop}%
\bibitem [{\citenamefont {Lee}(2015)}]{Lee2015}%
  \BibitemOpen
  \bibfield  {author} {\bibinfo {author} {\bibfnamefont {C.~F.}\ \bibnamefont
  {Lee}},\ }\href {http://arXiv.org/abs/1503.08674} {\bibfield  {journal}
  {\bibinfo  {journal} {arXiv:1503.08674}\ } (\bibinfo {year}
  {2015})}\BibitemShut {NoStop}%
\bibitem [{\citenamefont {Huber}\ and\ \citenamefont {Kim}(1996)}]{Huber1996}%
  \BibitemOpen
  \bibfield  {author} {\bibinfo {author} {\bibfnamefont {G.~A.}\ \bibnamefont
  {Huber}}\ and\ \bibinfo {author} {\bibfnamefont {S.}~\bibnamefont {Kim}},\
  }\href {\doibase 10.1016/S0006-3495(96)79552-8} {\bibfield  {journal}
  {\bibinfo  {journal} {Biophys. J.}\ }\textbf {\bibinfo {volume} {70}},\
  \bibinfo {pages} {97} (\bibinfo {year} {1996})}\BibitemShut {NoStop}%
\bibitem [{\citenamefont {Zhang}\ \emph {et~al.}(2010)\citenamefont {Zhang},
  \citenamefont {Jasnow},\ and\ \citenamefont {Zuckerman}}]{Zhang2010}%
  \BibitemOpen
  \bibfield  {author} {\bibinfo {author} {\bibfnamefont {B.~W.}\ \bibnamefont
  {Zhang}}, \bibinfo {author} {\bibfnamefont {D.}~\bibnamefont {Jasnow}}, \
  and\ \bibinfo {author} {\bibfnamefont {D.~M.}\ \bibnamefont {Zuckerman}},\
  }\href {\doibase 10.1063/1.3306345} {\bibfield  {journal} {\bibinfo
  {journal} {J. Chem. Phys.}\ }\textbf {\bibinfo {volume} {132}},\  (\bibinfo
  {year} {2010})}\BibitemShut {NoStop}%
\bibitem [{\citenamefont {Zhang}\ \emph {et~al.}(2007)\citenamefont {Zhang},
  \citenamefont {Jasnow},\ and\ \citenamefont {Zuckerman}}]{Zhang2007}%
  \BibitemOpen
  \bibfield  {author} {\bibinfo {author} {\bibfnamefont {B.~W.}\ \bibnamefont
  {Zhang}}, \bibinfo {author} {\bibfnamefont {D.}~\bibnamefont {Jasnow}}, \
  and\ \bibinfo {author} {\bibfnamefont {D.~M.}\ \bibnamefont {Zuckerman}},\
  }\href {\doibase 10.1073/pnas.0706349104} {\bibfield  {journal} {\bibinfo
  {journal} {Proc. Natl. Acad. Sci. U. S. A.}\ }\textbf {\bibinfo {volume}
  {104}},\ \bibinfo {pages} {18043} (\bibinfo {year} {2007})}\BibitemShut
  {NoStop}%
\bibitem [{Note1()}]{Note1}%
  \BibitemOpen
  \bibinfo {note} {We use NVT simulations rather than NPT (as would be typical
  in an equilibrium investigation of CNT) because the relevant experiments
  correspond to the NVT ensemble (\protect \textit {e.g.}\protect \xspace \cite
  {Palacci2013a}). Furthermore, defining the NPT ensemble requires a
  thermodynamic pressure. While such a pressure can be defined for the system
  under consideration \cite
  {Solon2015,Solon2015b,Takatori2014,marchetti2015structure,Takatori2015,Speck2015,Speck2015a,Winkler2015,Ginot2015},
  its use to implement NPT simulations remains as of yet untested. Our system
  sizes are large enough to avoid noticeable deviations from the thermodynamic
  limit which can occur in NVT simulations with extremely small system sizes
  \cite {Mayer1965,Thompson1984,Alder1962,Reguera2003}.}\BibitemShut {Stop}%
\bibitem [{\citenamefont {Auer}\ and\ \citenamefont
  {Frenkel}(2001)}]{Auer2001}%
  \BibitemOpen
  \bibfield  {author} {\bibinfo {author} {\bibfnamefont {S.}~\bibnamefont
  {Auer}}\ and\ \bibinfo {author} {\bibfnamefont {D.}~\bibnamefont {Frenkel}},\
  }\href {http://dx.doi.org/10.1038/35059035} {\bibfield  {journal} {\bibinfo
  {journal} {Nature}\ }\textbf {\bibinfo {volume} {409}},\ \bibinfo {pages}
  {1020} (\bibinfo {year} {2001})},\ \bibinfo {note}
  {10.1038/35059035}\BibitemShut {NoStop}%
\bibitem [{\citenamefont {Prestipino}\ \emph {et~al.}(2012)\citenamefont
  {Prestipino}, \citenamefont {Laio},\ and\ \citenamefont
  {Tosatti}}]{Prestipino2012}%
  \BibitemOpen
  \bibfield  {author} {\bibinfo {author} {\bibfnamefont {S.}~\bibnamefont
  {Prestipino}}, \bibinfo {author} {\bibfnamefont {A.}~\bibnamefont {Laio}}, \
  and\ \bibinfo {author} {\bibfnamefont {E.}~\bibnamefont {Tosatti}},\ }\href
  {\doibase 10.1103/PhysRevLett.108.225701} {\bibfield  {journal} {\bibinfo
  {journal} {Phys. Rev. Lett.}\ }\textbf {\bibinfo {volume} {108}},\ \bibinfo
  {pages} {225701} (\bibinfo {year} {2012})}\BibitemShut {NoStop}%
\bibitem [{\citenamefont {Loeffler}\ \emph {et~al.}(2013)\citenamefont
  {Loeffler}, \citenamefont {Henderson}, \citenamefont {Chen}, \citenamefont
  {Loeffler}, \citenamefont {Henderson},\ and\ \citenamefont
  {Chen}}]{Loeffler2013}%
  \BibitemOpen
  \bibfield  {author} {\bibinfo {author} {\bibfnamefont {T.~D.}\ \bibnamefont
  {Loeffler}}, \bibinfo {author} {\bibfnamefont {D.~E.}\ \bibnamefont
  {Henderson}}, \bibinfo {author} {\bibfnamefont {B.}~\bibnamefont {Chen}},
  \bibinfo {author} {\bibfnamefont {T.~D.}\ \bibnamefont {Loeffler}}, \bibinfo
  {author} {\bibfnamefont {D.~E.}\ \bibnamefont {Henderson}}, \ and\ \bibinfo
  {author} {\bibfnamefont {B.}~\bibnamefont {Chen}},\ }\href {\doibase
  10.1063/1.4766328} {\bibfield  {journal} {\bibinfo  {journal} {J. Chem.
  Phys.}\ }\textbf {\bibinfo {volume} {137}},\ \bibinfo {pages} {19430}
  (\bibinfo {year} {2013})}\BibitemShut {NoStop}%
\bibitem [{\citenamefont {Statt}\ \emph {et~al.}(2015)\citenamefont {Statt},
  \citenamefont {Virnau},\ and\ \citenamefont {Binder}}]{Statt2015}%
  \BibitemOpen
  \bibfield  {author} {\bibinfo {author} {\bibfnamefont {A.}~\bibnamefont
  {Statt}}, \bibinfo {author} {\bibfnamefont {P.}~\bibnamefont {Virnau}}, \
  and\ \bibinfo {author} {\bibfnamefont {K.}~\bibnamefont {Binder}},\ }\href
  {\doibase 10.1103/PhysRevLett.114.026101} {\bibfield  {journal} {\bibinfo
  {journal} {Phys. Rev. Lett.}\ }\textbf {\bibinfo {volume} {114}},\ \bibinfo
  {pages} {26101} (\bibinfo {year} {2015})}\BibitemShut {NoStop}%
\bibitem [{\citenamefont {Mcgraw}\ \emph {et~al.}(1997)\citenamefont {Mcgraw},
  \citenamefont {Laaksonen},\ and\ \citenamefont {Mcgraw}}]{Mcgraw1997}%
  \BibitemOpen
  \bibfield  {author} {\bibinfo {author} {\bibfnamefont {R.}~\bibnamefont
  {Mcgraw}}, \bibinfo {author} {\bibfnamefont {A.}~\bibnamefont {Laaksonen}}, \
  and\ \bibinfo {author} {\bibfnamefont {R.}~\bibnamefont {Mcgraw}},\ }\href
  {\doibase 10.1063/1.473527} {\bibfield  {journal} {\bibinfo  {journal} {J.
  Chem. Phys.}\ }\textbf {\bibinfo {volume} {106}},\ \bibinfo {pages} {5284}
  (\bibinfo {year} {1997})}\BibitemShut {NoStop}%
\bibitem [{Note2()}]{Note2}%
  \BibitemOpen
  \bibinfo {note} {While the theoretical CSD does predict a power law
  (represented by the logarithmic correction to $G(n)$ in Eq. \ref
  {eq:kinetic_free_energy_2}), it is suppressed in the region of interest and
  has exponent ${\raise .17ex\hbox {$\scriptstyle \protect \mathtt {\sim }$}}
  0.59$ instead of $2$.}\BibitemShut {Stop}%
\bibitem [{\citenamefont {Levis}\ and\ \citenamefont
  {Berthier}(2014)}]{Levis2014}%
  \BibitemOpen
  \bibfield  {author} {\bibinfo {author} {\bibfnamefont {D.}~\bibnamefont
  {Levis}}\ and\ \bibinfo {author} {\bibfnamefont {L.}~\bibnamefont
  {Berthier}},\ }\href {\doibase 10.1103/PhysRevE.89.062301} {\bibfield
  {journal} {\bibinfo  {journal} {Phys. Rev. E}\ }\textbf {\bibinfo {volume}
  {89}},\ \bibinfo {pages} {062301} (\bibinfo {year} {2014})}\BibitemShut
  {NoStop}%
\bibitem [{\citenamefont {Reguera}\ and\ \citenamefont
  {Rub\'{i}}(2001)}]{Reguera2001}%
  \BibitemOpen
  \bibfield  {author} {\bibinfo {author} {\bibfnamefont {D.}~\bibnamefont
  {Reguera}}\ and\ \bibinfo {author} {\bibfnamefont {J.}~\bibnamefont
  {Rub\'{i}}},\ }\href {\doibase 10.1063/1.1405122} {\bibfield  {journal}
  {\bibinfo  {journal} {J. Chem. Phys.}\ }\textbf {\bibinfo {volume} {115}},\
  \bibinfo {pages} {7100} (\bibinfo {year} {2001})}\BibitemShut {NoStop}%
\bibitem [{\citenamefont {Kiang}\ \emph {et~al.}(1971)\citenamefont {Kiang},
  \citenamefont {Stauffer}, \citenamefont {Walker}, \citenamefont {Puri},
  \citenamefont {Wise},\ and\ \citenamefont {Patterson}}]{Kiang1971}%
  \BibitemOpen
  \bibfield  {author} {\bibinfo {author} {\bibfnamefont {C.}~\bibnamefont
  {Kiang}}, \bibinfo {author} {\bibfnamefont {D.}~\bibnamefont {Stauffer}},
  \bibinfo {author} {\bibfnamefont {G.}~\bibnamefont {Walker}}, \bibinfo
  {author} {\bibfnamefont {O.}~\bibnamefont {Puri}}, \bibinfo {author}
  {\bibfnamefont {J.~J.}\ \bibnamefont {Wise}}, \ and\ \bibinfo {author}
  {\bibfnamefont {E.}~\bibnamefont {Patterson}},\ }\href@noop {} {\bibfield
  {journal} {\bibinfo  {journal} {J. Atmos. Sci.}\ }\textbf {\bibinfo {volume}
  {28}},\ \bibinfo {pages} {1222} (\bibinfo {year} {1971})}\BibitemShut
  {NoStop}%
\bibitem [{\citenamefont {Fisher}(1967)}]{Fisher1967}%
  \BibitemOpen
  \bibfield  {author} {\bibinfo {author} {\bibfnamefont {M.~E.}\ \bibnamefont
  {Fisher}},\ }\href@noop {} {\bibfield  {journal} {\bibinfo  {journal}
  {Physics}\ }\textbf {\bibinfo {volume} {3}},\ \bibinfo {pages} {255}
  (\bibinfo {year} {1967})}\BibitemShut {NoStop}%
\bibitem [{Note3()}]{Note3}%
  \BibitemOpen
  \bibinfo {note} {While completing this manuscript, we learned of another
  concurrent study which computationally measures ABP nucleation rates \cite
  {Richard2016}. By analyzing simulation trajectories, they conclude that
  cluster size and cluster polarization (the extent to which particles on the
  rim of a cluster point inwards) are important collective coordinates for
  describing transition rates. These observations are consistent with the
  kinetic theory described here.}\BibitemShut {Stop}%
\bibitem [{\citenamefont {Edelsbrunner}\ \emph {et~al.}(1983)\citenamefont
  {Edelsbrunner}, \citenamefont {Kirkpatrick},\ and\ \citenamefont
  {Seidel}}]{Edelsbrunner1983}%
  \BibitemOpen
  \bibfield  {author} {\bibinfo {author} {\bibfnamefont {H.}~\bibnamefont
  {Edelsbrunner}}, \bibinfo {author} {\bibfnamefont {D.~G.}\ \bibnamefont
  {Kirkpatrick}}, \ and\ \bibinfo {author} {\bibfnamefont {R.}~\bibnamefont
  {Seidel}},\ }\href {\doibase 10.1109/TIT.1983.1056714} {\bibfield  {journal}
  {\bibinfo  {journal} {Information Theory, IEEE Transactions on}\ }\textbf
  {\bibinfo {volume} {29}},\ \bibinfo {pages} {551} (\bibinfo {year}
  {1983})}\BibitemShut {NoStop}%
\bibitem [{\citenamefont {Mayer}\ and\ \citenamefont {Wood}(1965)}]{Mayer1965}%
  \BibitemOpen
  \bibfield  {author} {\bibinfo {author} {\bibfnamefont {J.~E.}\ \bibnamefont
  {Mayer}}\ and\ \bibinfo {author} {\bibfnamefont {W.~W.}\ \bibnamefont
  {Wood}},\ }\href {\doibase 10.1063/1.1695931} {\bibfield  {journal} {\bibinfo
   {journal} {J. Chem. Phys.}\ }\textbf {\bibinfo {volume} {42}},\ \bibinfo
  {pages} {4268} (\bibinfo {year} {1965})}\BibitemShut {NoStop}%
\bibitem [{\citenamefont {Thompson}\ \emph {et~al.}(1984)\citenamefont
  {Thompson}, \citenamefont {Gubbins}, \citenamefont {Walton}, \citenamefont
  {Chantry},\ and\ \citenamefont {Rowlinson}}]{Thompson1984}%
  \BibitemOpen
  \bibfield  {author} {\bibinfo {author} {\bibfnamefont {S.~M.}\ \bibnamefont
  {Thompson}}, \bibinfo {author} {\bibfnamefont {K.~E.}\ \bibnamefont
  {Gubbins}}, \bibinfo {author} {\bibfnamefont {J.~P. R.~B.}\ \bibnamefont
  {Walton}}, \bibinfo {author} {\bibfnamefont {R.~A.~R.}\ \bibnamefont
  {Chantry}}, \ and\ \bibinfo {author} {\bibfnamefont {J.~S.}\ \bibnamefont
  {Rowlinson}},\ }\href {\doibase 10.1063/1.447358} {\bibfield  {journal}
  {\bibinfo  {journal} {J. Chem. Phys.}\ }\textbf {\bibinfo {volume} {81}},\
  \bibinfo {pages} {530} (\bibinfo {year} {1984})}\BibitemShut {NoStop}%
\bibitem [{\citenamefont {Alder}\ and\ \citenamefont
  {Wainwright}(1962)}]{Alder1962}%
  \BibitemOpen
  \bibfield  {author} {\bibinfo {author} {\bibfnamefont {B.~J.}\ \bibnamefont
  {Alder}}\ and\ \bibinfo {author} {\bibfnamefont {T.~E.}\ \bibnamefont
  {Wainwright}},\ }\href {\doibase 10.1103/PhysRev.127.359} {\bibfield
  {journal} {\bibinfo  {journal} {Phys. Rev.}\ }\textbf {\bibinfo {volume}
  {127}},\ \bibinfo {pages} {359} (\bibinfo {year} {1962})}\BibitemShut
  {NoStop}%
\bibitem [{\citenamefont {Reguera}\ \emph {et~al.}(2003)\citenamefont
  {Reguera}, \citenamefont {Bowles}, \citenamefont {Djikaev},\ and\
  \citenamefont {Reiss}}]{Reguera2003}%
  \BibitemOpen
  \bibfield  {author} {\bibinfo {author} {\bibfnamefont {D.}~\bibnamefont
  {Reguera}}, \bibinfo {author} {\bibfnamefont {R.~K.}\ \bibnamefont {Bowles}},
  \bibinfo {author} {\bibfnamefont {Y.}~\bibnamefont {Djikaev}}, \ and\
  \bibinfo {author} {\bibfnamefont {H.}~\bibnamefont {Reiss}},\ }\href
  {\doibase 10.1063/1.1524192} {\bibfield  {journal} {\bibinfo  {journal} {J.
  Chem. Phys.}\ }\textbf {\bibinfo {volume} {118}},\ \bibinfo {pages} {340}
  (\bibinfo {year} {2003})}\BibitemShut {NoStop}%
\end{thebibliography}%

\end{document}